\documentclass[aps,pra,twocolumn,superscriptaddress]{revtex4-2}
\usepackage{natbib}
\usepackage{graphicx}
\usepackage{amsmath}
\usepackage{mathtools}
\usepackage{amssymb}
\usepackage{amsmath}
\usepackage[colorlinks,urlcolor=blue,citecolor=blue,linkcolor=blue]{hyperref}
\usepackage{url}
 
\begin{document}

\title{Finite-frequency normal and superfluid drag effects\\
in two-component atomic Bose-Einstein condensates}

\author{Azat F. Aminov}
\email{afaminov@hse.ru}
\affiliation{National Research University Higher School of Economics, 109028 Moscow, Russia}
\affiliation{Institute of Microelectronics Technology and High Purity Materials, Russian Academy of Sciences, Chernogolovka 142432, Russia}

\author{Alexey A. Sokolik}
\email{asokolik@hse.ru}
\affiliation{Institute for Spectroscopy, Russian Academy of Sciences, 142190 Troitsk, Moscow, Russia}
\affiliation{National Research University Higher School of Economics, 109028 Moscow, Russia}

\author{Yurii E. Lozovik}
\affiliation{Institute for Spectroscopy, Russian Academy of Sciences, 142190 Troitsk, Moscow, Russia}
\affiliation{National Research University Higher School of Economics, 109028 Moscow, Russia}

\begin{abstract}
Two-component systems consisting of mutually interacting particles can demonstrate both intracomponent transport effects and intercomponent entrainment (or drag) effects. In the presence of superfluidity, the intracomponent transport is characterized by dissipative conductivity and superfluid weight in the framework of two-fluid model, and intercomponent entrainment gives rise to normal and nondissipative drag effects. We present unified treatment of all these effects for spatially homogeneous two-component atomic Bose-Einstein condensates based on the Bogoliubov theory, focusing specifically on the drag effects. Calculating finite-frequency intra- and intercomponent conductivities with taking into account quasiparticle damping, we derive and numerically check analytical Drude-like approximations applicable at low frequencies, and Lorentz-like approximations applicable at higher frequencies in vicinity of the resonant energy of spin-to-density Bogoliubov quasiparticle conversion. As possible physical realizations of two-component atomic systems, we consider three-dimensional Bose-Bose mixtures and closely spaced two-layered systems of magnetic dipolar atoms.
\end{abstract}

\maketitle

\section{Introduction}

Understanding of many-body phenomena in ultracold atomic gases helps to shed light on the properties of condensed matter systems. For example, studying Bose-Einstein condensation (BEC) of ultracold atomic gases provides deeper insight into physics of superconductors, superfluids, and strongly correlated systems \cite{Chien_2015,Proukakis_2017,Bloch2012,Mistakidis}. One of such phenomena, which might occur in semiconducting, superconducting, and ultracold atomic systems, is drag effect, the transport phenomenon which reveals both single-particle and many-body physics.

The Coulomb drag effect in closely spaced two-layer systems, which is caused by frictional entrainment of particles in one layer in response to a current in the other layer, is extensively studied in solid-state electronic systems \cite{Narozhny2016}. Experimentally, this effect is detected by measuring nonlocal transresistance between layers. In superfluid or superconducting two-component systems, a non-dissipative counterpart of the drag, or Andreev-Bashkin effect, can also emerge, when superfluid or superconducting components of the constituents entrain each other without dissipation. This effect was predicted for $^{3}$He-$^4$He mixtures \cite{AB}, superconducting systems \cite{Duan1993,Yerin_2022,Yerin_2023}, ultracold atomic gases \cite{Tanatar_1996}, superfluid mixtures of nucleons in the cores of neutron stars \cite{Alpar1984}, and for superconducting layers interacting with polaritons \cite{Aminov_PRB}. 

Both Coulomb and Andreev-Bashkin drag effects are conventionally studied in the DC regime (at $\omega=0$). Recently there appeared an interest in studying the AC ($\omega>0$) drag effect \cite{Sekino_2023}: an alternating force at nonzero frequency acts upon one component, and the alternating current of the other component is detected. This effect is described by the conductivity matrix $\sigma_{ij}(\omega)$ resolved over the components $i,j=a,b$. The key feature of AC drag effects in a superfluid system is interplay of dissipative and non-dissipative current responses, when the Coulomb drag and Andreev-Bashkin effects in their pure DC form can be extracted from analysis of the low-frequency limit of AC drag conductivity $\sigma_{ab}(\omega)$.

\begin{figure}[t]
\begin{center}
\includegraphics[width=1\columnwidth]{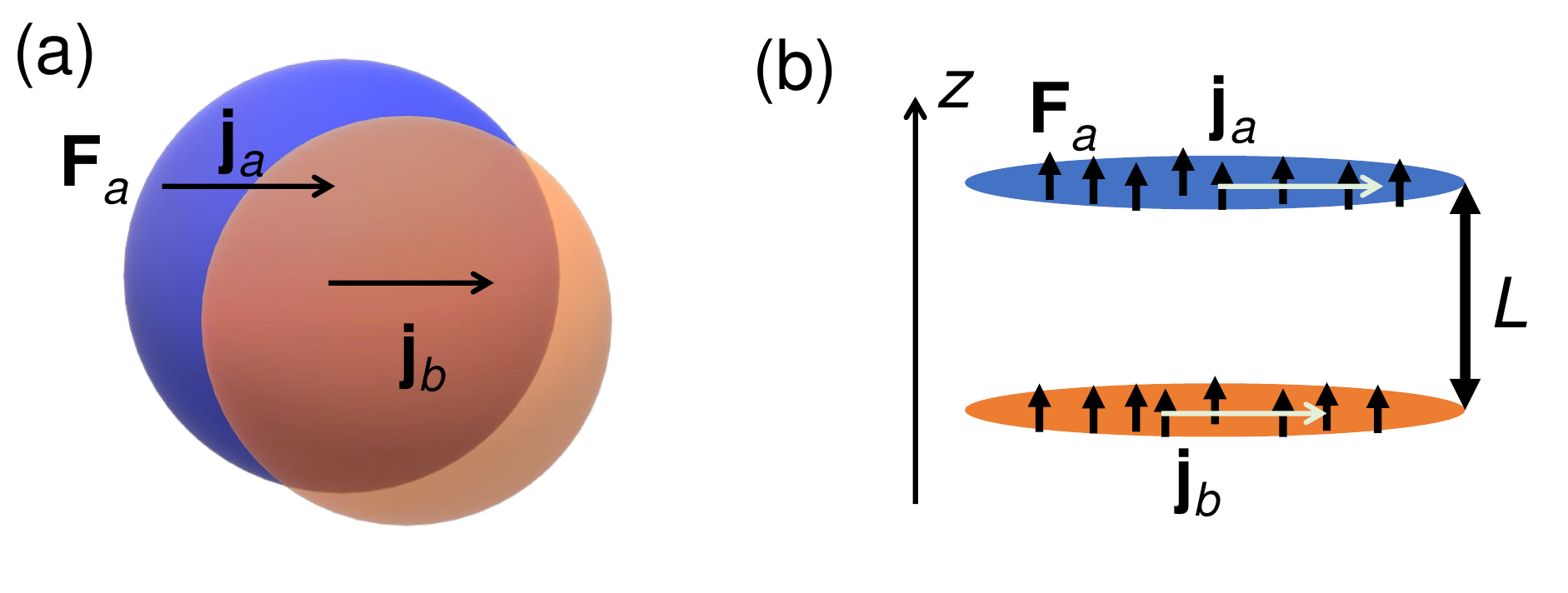}
\end{center}
\caption{Schematic depiction of systems considered. (a) 3D atomic mixture. Alternating force $\mathbf{F}_{a}$ imposed on the constituent $a$ gives rise to both intracomponent $\mathbf{j}_{a}\sim\sigma_{aa}\mathbf{F}_a$ and intercomponent $\mathbf{j}_{b}\sim \sigma_{ab}\mathbf{F}_a$ response currents. (b) Dipolar atomic quasi-2D system. Atoms in two pancake-like Bose-condensed clouds have their dipole moments aligned with the $z$ axis, and the drag effects are induced by long-range interaction across the interlayer distance $L$.}
\label{fig:Scheme}
\end{figure}

In this paper we calculate, using many-body theory, the AC mass conductivities $\sigma_{ij}(\omega)$ of a two-component atomic BEC at nonzero temperature, considered as a homogeneous 3D mixture [see Fig.~\ref{fig:Scheme}(a)]. We analyze both intracomponent conductivities (or \emph{intraconductivities}) $\sigma_{aa}$, $\sigma_{bb}$, which characterize normal and superfluid responses of each component, and intercomponent conductivity (or \emph{transconductivity}) $\sigma_{ab}$, which is responsible for normal and superfluid drag effects. In contrast to Ref.~\cite{Sekino_2023}, we consider generally non-symmetric two-component system with different masses and densities of constituents, and assume nonzero damping $\gamma$ of the Bogoliubov excitations, which can be caused by interatomic interactions with thermally excited quasiparticles \cite{Pitaevskii_1997,Chung_2009} or by scattering on impurities or disorder \cite{Giorgini_1994,Lopatin_2002,Gaul_2011,Muller_2012}. The external disorder potential can arise due to experimental imperfections or can be introduced intentionally \cite{Nagler_2020,Zhou_2014}.

Besides three-dimensional mixtures, we study spatially separated magnetic dipolar atomic gases [see Fig.~\ref{fig:Scheme}(b)], where the interlayer drag can appear due to long-range dipole-dipole interaction. Such systems are gaining popularity nowadays: BEC of dipolar atomic gases was realised in recent experiments \cite{Politi_2022}, and mutual friction (i.e. normal drag effect) in a non-condensed phase was detected in the two-layered geometry \cite{DuLi_2023}.

We calculate and analyze frequency dependencies of dissipative (or real) and non-dissipative (or imaginary) parts of the conductivities $\sigma_{ij}(\omega)$, which characterize the current responses in phase and with the $\pi/2$ phase shift with respect to the driving force, respectively. At large enough temperatures the dissipative and non-dissipative response currents may be of the same order, leading to non-trivial phase shifts (besides 0 and $\pi/2$) between currents and driving forces. At low frequencies both intra- and transconductivities $\sigma_{ij}(\omega)$ are well approximated analytically by a kind of two-fluid Drude model \cite{Brorson_1996,Yang_2018} with a mixture of nondissipative response (giving rise to superfluidity of each component and Andreev-Bashkin effect between the components) and dissipative response caused by quasiparticle decay (which gives rise to a normal conductivity of each component and normal drag between the components). At higher frequencies of the order of atomic chemical potentials, $\sigma_{ij}(\omega)$ in certain conditions can reveal the Lorentz-type resonance originating from interconversion between spin and density Bogoliubov quasiparticles, which is also analytically approximated. For dipolar atoms with interlayer interaction, we predict similar behavior of transconductivity, although the resonance frequency may be tuned by changing the interlayer distance. 

The paper is structured as follows. In Sec. \ref{sec:Theory} the outline of the theory is presented, providing the general expressions for conductivity calculations and parameters of the atomic systems we consider. Then in Sec. \ref{sec:Analytic} we derive analytic approximations for AC conductivities in the Drude (low-frequency) and Lorentz (high-frequency) regimes, followed by Sec. \ref{sec:NumCalc}, where the results of numerical calculations are presented and compared with the analytical approximations. Sec. \ref{sec:Discussion} concludes the paper with discussion. Appendices \ref{app:BT}, \ref{app:Details}, \ref{app:Delta}, \ref{app:d-dint} present details of calculations.

\section{Theory}\label{sec:Theory}
\subsection{Intra- and transconductivities}

Superfluid, dissipative and drag transport effects in a two-component system are characterized by AC intraconductivities $\sigma_{aa}(\omega)$, $\sigma_{bb}(\omega)$, and transconductivity $\sigma_{ab}(\omega)$, which relate the force $\mathbf{F}_{j}e^{-i\omega t}$ imposed on the component $j$ to the current density (or flux density of particles) induced in the component $i$:
\begin{equation}
    \mathbf{j}_{i}(t)=\sigma_{ij}(\omega) \mathbf{F}_{j}e^{-i\omega t}.
\end{equation}
Such conductivities have dimensionality of $\hbar^{-1}\mbox{cm}^{-1}$ ($\hbar^{-1}$) for 3D (2D) system. In experiments on ultracold atomic gases, they can be determined by measuring velocities and coordinates of atoms using time-of-flight expansion imaging \cite{Lee_2018} or temperature change due to dissipation-induced heating \cite{Llorente_Garc_a_2013}. 

In the linear response theory, the AC conductivities can be related to the retarded correlation functions of currents \cite{Romito2020,Sekino_2023}
\begin{equation}
\sigma_{ij}(\omega) = \frac{i}{\omega}\left[\frac{\delta_{ij}n_{i}}{m_{i}} + \lim_{q\to0} \chi^{\mathrm{T}}_{ij}(q,\omega) \right].\label{basis}
\end{equation}
Here the first term is diamagnetic response present only in the intracomponent channel $i=j$, with $n_i$ and $m_i$ being the atomic density and mass of the $i$th component; $\chi^{\mathrm{T}}_{ij}$ is transverse part of the retarded paramagnetic current response tensor. In Matsubara representation at nonzero temperature $T$, this tensor is given by 
\begin{equation}
\chi_{ij}^{\nu \eta}(\mathbf{q},i\omega)=-\frac1A\int\limits_0^{1/T}d\tau\:e^{i\omega\tau}\left\langle T_\tau j_{i}^{\nu}(\mathbf{q},\tau)j^{\eta}_{j}(-\mathbf{q},0)\right\rangle,\label{chi1}
\end{equation}
where $j^{\nu}_{i}(\mathbf{q},\tau)$ is the Heisenberg-evolved (in imaginary time) operator of the $\mathbf{q}$th spatial harmonic of the $i$th component current density along the axis $\nu$, $A$ is the system volume (area) in the case of 3D (2D) geometry, and hereafter $\hbar=1$ is assumed in the formulas. The retarded correlation function of currents entering Eq.~(\ref{basis}) can be obtained from the Matsubara one (\ref{chi1}) by taking transverse tensorial part over $\nu$, $\eta$, and performing analytic continuation $i\omega\rightarrow\omega+i0$ from the upper half of the complex plane. Since we are interested in response of currents on a homogeneous force $\mathbf{F}_{i}$, we take $\mathbf{q}=0$ in Eq.~(\ref{chi1}). Note, however, that the $\omega\to 0$ limit has to be taken carefully in DC conductivity calculations. It can be shown \cite{Scalapino}, that in order to correctly calculate the superfluid drag density \cite{Fil2005} in a system without quasiparticle damping, the DC limit $\omega\to0$ has to be taken before the $\mathbf{q}\to0$ limit, although in the presence of damping these two limits commute, which will be used below.

The induced current can be divided into the in-phase (with respect to the driving force) part, which is responsible for dissipation, and $\pi/2$ phase-delayed part, which is non-dissipative. The latter can be additionally divided into the diamagnetic and paramagnetic contributions. By this reason, imaginary part of each conductivity 
\begin{equation}
\sigma_{ij}^{\mathrm{s}}(\omega)\equiv \mathrm{Im}\,\sigma_{ij}(\omega) = \frac{\delta_{ij} n_{i}}{m_{i}\omega}+\sigma_{ij}^{\mathrm{sp}}(\omega),\label{sigma_s}
\end{equation}
which consists of dia- ($\delta_{ij} n_{i}/m_{i}$) and paramagnetic ($\sigma_{ij}^{\mathrm{sp}}$) parts, will be referred to as \emph{superfluid} conductivity, and the real part 
\begin{equation}
\sigma_{ij}^{\mathrm{n}}(\omega)\equiv \mathrm{Re}\,\sigma_{ij}(\omega)
\end{equation}
will be referred to as \emph{normal} conductivity. Note that the distinction between superfluid and normal responses is strictly defined only in the DC limit $\omega=0$ \cite{Griffin,Brorson_1996,Yang_2018}, where the $1/\omega$ singularity of $\sigma^\mathrm{s}_{ij}$ indicates superfluidity in both inter- and intracomponent channels. In particular, the theory of DC superfluid drag \cite{Fil2005} deals with the superfluid drag mass density
\begin{equation}
    \rho_{\mathrm{dr}} = m_{a}m_{b} \lim_{\omega\to 0} \omega\sigma_{ab}^{\mathrm{s}}(\omega),\label{rho_dr}
\end{equation}
and the intracomponent superfluid mass density is related to the low-frequency divergence of the intraconductivity 
\begin{equation}
    \rho_i^\mathrm{s} = m_i^2\lim_{\omega\to 0} \omega\sigma_{ii}^{\mathrm{s}}(\omega).\label{rho_s}
\end{equation}
In contrast, $\sigma^\mathrm{n}_{ij}(\omega)$ tends to a constant in the limit $\omega\rightarrow0$, and its intracomponent part $\sigma^\mathrm{n}_{ii}(0)$ provides dissipative DC conductivity, while the intercomponent part $\sigma^\mathrm{n}_{ab}(0)$ is related to the DC drag coefficient, or transresistivity $\sigma_{ab}^{\mathrm{n}}(0)/[\sigma_{aa}^{\mathrm{n}}(0)\sigma_{bb}^{\mathrm{n}}(0) - \sigma_{ab}^{\mathrm{n}}(0)^{2}]$, which is usually measured in drag experiments \cite{Narozhny2016}. 

At $\omega>0$, the strict distinction between superfluid and normal responses becomes elusive because even in normal systems both $\sigma_{ij}^{\mathrm{n}}$ and $\sigma_{ij}^{\mathrm{s}}$ are finite and nonzero. Detecting the dissipative response of a normal current against the superfluid background is harder task for Bose systems than for conventional s-wave superconductors, where the dissipative part of conductivity is suppressed at $\omega < 2\Delta$ \cite{Mattis_1958,Sekino_2022}. In contrast, Bose-condensed systems lack gap in the quasiparticle spectrum, so the normal conductivity is generally nonzero at any $\omega>0$ \cite{Sekino_2022}.
This is why analysis of AC conductivities $\sigma_{ij}(\omega)$ provides unified and more detailed information about both the superfluid entraintment (Andreev-Bashkin) effect and normal drag effect, as well as about normal (dissipative) and superfluid (nondissipative) responses of each component, than conventional DC calculations commonly accepted in the theories of drag and superfluidity.

As specific examples of 3D mixtures [Fig. \ref{fig:Scheme}(a)], we consider spinor atomic BECs: the symmetric mixtures of $^{87}$Rb and $^{23}$Na in atomic states $F=1, m_{F}=\pm 1$ \cite{Fava_2018}, and  the non-symmetric mixture of $^{39}$K in the states $F=1, m_F= 1$ and $F=1, m_F= 0$ \cite{Cabrera_2018}; here $m_F$ are magnetic sublevels of hyperfine state with total angular momentum $F$. Besides, we consider the mixture of atoms with different masses, $^{174}$Yb-$^{133}$Cs \cite{Wilson_2021}. As the spatially separated quasi-2D dipolar system [Fig. \ref{fig:Scheme}(b)], we consider pairs of parallel clouds of either $^{52}$Cr or $^{168}$Er atoms with the long-range magnetic dipole interaction \cite{Young_S__2012}.

The realistic parameters used in numerical calculations are listed in the Table~\ref{table:1}. Each intra- or intercomponent interaction constant $g_{ij} = 2\pi a^\mathrm{s}_{ij} (1/m_{i} + 1/m_{j})$ is related to the s-wave scattering length $a^\mathrm{s}_{ij}$, and we neglect the processes which permit population transfer between different magnetic sublevels $m_F$ of the state with total angular momentum $F$. In the case of spatially separated dipolar atomic clouds, we take into account both s-wave scattering and dipole-dipole interaction within each cloud, and only the dipole interaction between atoms from different clouds (see details in Appendix~\ref{app:d-dint}). In accordance with the recent experiment \cite{DuLi_2023}, the thickness of both clouds is assumed to be $w_{z}=20\,\mbox{nm}$, and the distance between clouds is $L=60\,\mbox{nm}$.

In this paper we consider systems with large condensate fraction, when the temperature is much lower than the BEC critical temperatures of both constituents, $T\ll T_\mathrm{c}^i$, but nonzero, since we are interested in the normal drag effect as well. For 3D homogeneous atomic gases, the condensate densities $n_i^0$ are found with taking into account their thermal and quantum depletions from the system of equations $n_i^0(T)=n_{i}-n_{i}^{\mathrm{nc}}(n_a^0,n_b^0,T)$, $i=a,b$, where $n_{i}^{\mathrm{nc}}$ is a density of non-condensed fraction given by Eq.~(\ref{non-cond}), $n_{i}$ is the total density of the $i$ component, which is assumed to be temperature-independent and estimated from the experimental critical temperature as $n_{i}=\zeta (3/2) \left(m_{i}T_\mathrm{c}^i/2\pi\right)^{3/2}$. The sums of chemical potentials $\sum_{i}\mu_{i}=\sum_{i}g_{ii}n^{0}_{i}$ listed in Table~\ref{table:1}, which provide characteristic energy scales of excitation energies, are taken at zero temperature.

\begin{table}[t]
\centering
\begin{tabular}{ccccc}
\multicolumn{5}{c}{Non-dipolar atoms} \\
\hline
\hline
Parameter & $^{87}$Rb & $^{23}$Na & $^{39}$K & $^{174}$Yb-$^{133}$Cs\\
\hline
$a^\mathrm{s}_{ii} (a_{0})$ & 100  & 55 & 30, 100 & 105, 150 \\
$a^\mathrm{s}_{ab} (a_{0})$ & 95  & 51 & $-50$ & $-75$ \\
$T_{\mathrm{c}}$ (nK) & 170  & 1000 & 150 & 460, 200\\
$\sum_{i} \mu_{i}/2\pi$ (kHz) & 2.7 & 10.8 & 1 & 12.7 \\
\hline
\hline
\end{tabular}

\begin{tabular}{cc c}
\multicolumn{3}{c}{Dipolar atoms} \\
\hline
\hline
Parameter & $^{52}$Cr & $^{168}$Er \\
\hline
$d_{i} (\mu_{\mathrm{B}})$ & 6  & 7 \\
$a^\mathrm{s}_{ii} (a_{0})$ & 103  & 137 \\
$T_{\mathrm{c}}$ (nK) & 700 & 410 \\
\hline
\hline
\end{tabular}
\caption{Upper table: Parameters for 3D spinor mixtures, namely intra- $a^\mathrm{s}_{ii}$ and intercomponent $a^\mathrm{s}_{ab}$ scattering lengths in the units of the Bohr radius $a_{0}\approx0.529\,\mbox{\AA}$, critical temperatures $T_\mathrm{c}$, and sums of chemical potentials $\mu_i$ of the components at $T=0$ (separated by a comma for non-symmetric mixtures). Lower table: parameters for dipolar atoms including magnetic dipole moments $d_{i}$ in Bohr magneton $\mu_{\mathrm{B}}$ units.}
\label{table:1}
\end{table}

\subsection{Current response function}

In order to calculate the current response function (\ref{chi1}), we use diagrammatic technique to express it in the single-loop approximation through the intra- and intercomponent matrix Green functions $\hat{G}_{ij}$. Explicit formulas for the Green functions are provided in Appendix~\ref{app:BT}, and calculation details for the current response are given in Appendix~\ref{app:Details}.

\begin{figure}[t]
\begin{center}
\includegraphics[width=0.93\columnwidth]{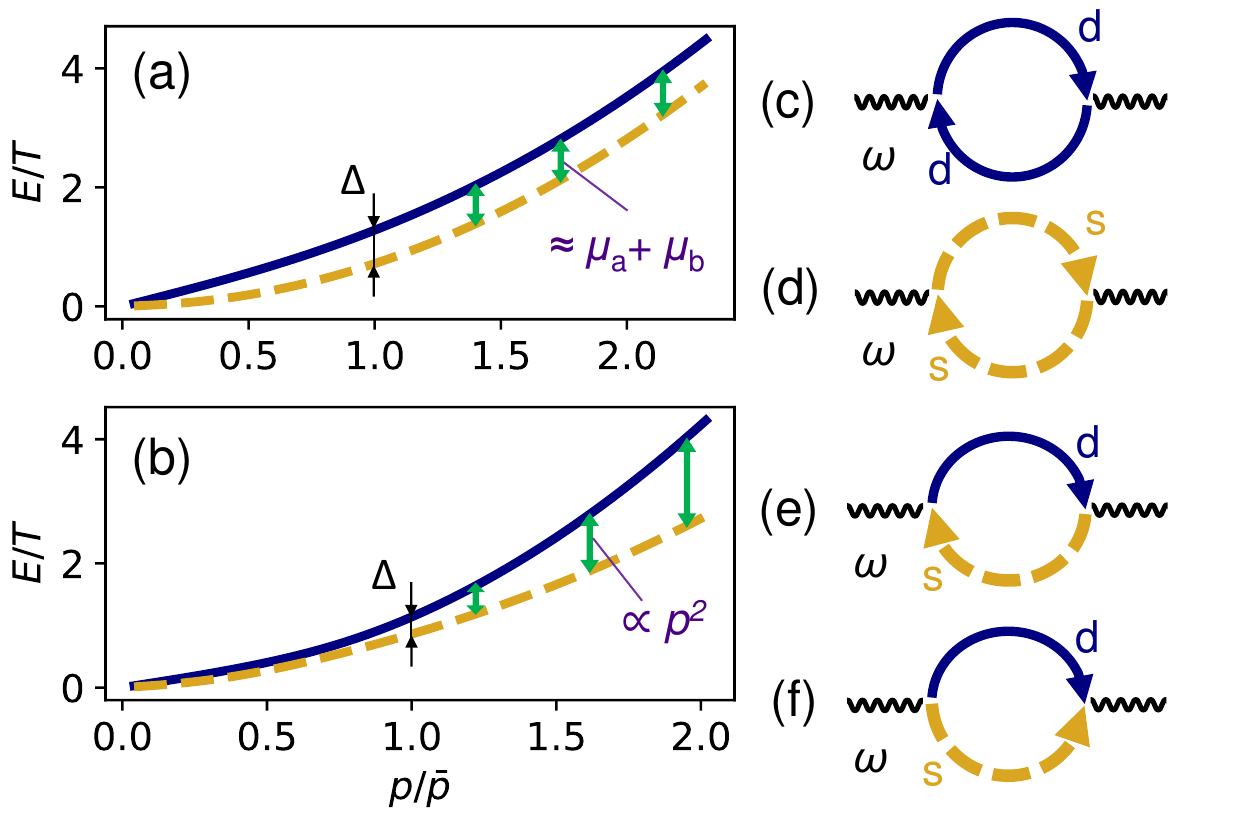}
\end{center}
\caption{Left panels: Bogoliubov quasiparticle dispersions $E_\mathrm{d,s}(p)$ in the cases of close (a) and distant (b) atomic masses. Green arrows indicate energy differences $E_\mathrm{d}(p)-E_\mathrm{s}(p)$ at the relevant momenta $p\sim\bar{p}$. Right panels (c-f): excitations of pairs of the Bogoliubov quasiparticles contributing to conductivities.}
\label{fig:Edif}
\end{figure}

It is known \cite{Pu_1998, Fil2005}, that in a two-component BEC two types of quasiparticles emerge, which correspond to density and spin collective modes, with dispersions $E_{\mathrm{d}}(q)$ and $E_{\mathrm{s}}(q)$, respectively [see Fig.~\ref{fig:Edif}(a-b)]. Calculating the total current response function $\chi_{ij}^{\mu\nu}$, we express it through the response functions $S(E_{\alpha},E_{\beta})$ resolved over the quasiparticle branches $\alpha, \beta = \mathrm{d, s}$ and weighted with Bogoliubov coefficients. For the transverse part of the current response tensor (\ref{chi1}), we obtain
\begin{align}
\chi^{\mathrm{T}}_{ij}(q=0,i\omega)=\sum_{\mathbf{p}}\frac{p^2}{2Adm_{i}m_{j}}\sum_{\alpha_1 \alpha_2s_1s_2}s_1 s_2&\nonumber\\
\times \left(u_{i \alpha_1}^{s_1} u_{i \alpha_2}^{s_2}-u_{i \alpha_1}^{-s_1} u_{i \alpha_2}^{-s_2}\right) \left(u_{j \alpha_1}^{s_1} u_{j \alpha_2}^{s_2}-u_{j \alpha_1}^{-s_1} u_{j \alpha_2}^{-s_2}\right)&\nonumber\\
\times S(s_1E_{\alpha_{1}},s_2E_{\alpha_{2}}),&\label{chiSF}
\end{align}
where $u_{i\alpha}^+$ and $u_{i\alpha}^-$ play the role of $u$ and $v$ Bogoliubov coefficients for the $i$th component and the $\alpha$th branch. The sums are taken over $d$-dimensional momentum $\mathbf{p}$, positive and negative energy indices $s_{1,2}=\pm$, and quasiparticle branches $\alpha_{1,2}=\mathrm{d,s}$. The response function
\begin{equation}
    S(E_{\alpha},E_{\beta}) = - T \sum_{i\omega_{n}} \frac{1}{(i\omega_{n}-E_{\alpha}+i\omega)(i\omega_{n}-E_{\beta})}\label{S_clean}
\end{equation}
corresponds to the loop-diagram constructed from two Matsubara Green functions $1/(i\omega_{n}- E_{\mathrm{\alpha}})$ of Bogoliubov quasiparticles, which have infinite lifetime. Since we aim to analyze both normal and superfluid drag effects, we ought to account for their non-zero damping $\gamma$ by replacing the quasiparticle Green functions $1/(i\omega_{n}- E_{\mathrm{\alpha}})$ with the broadened ones $\int\,dx\:\rho_{\alpha}(x)/(i\omega_{n}-x)$, where $\rho_{\alpha}(x)=(\gamma/\pi) [(x-E_{\alpha})^{2}+\gamma^{2}]^{-1}$ is the Lorentzian spectral function.  For simplicity of the forthcoming analytical calculations, we assume $\gamma$ to be momentum- and energy-independent and to be the same for both spin and density modes. In this approximation the sum over Matsubara frequencies in Eq.~(\ref{S_clean}) can be taken analytically:
\begin{equation}
    S(E_{\alpha},E_{\beta})= \int  dx dx^{\prime}\,\rho_{\alpha}(x) \rho_{\beta}(x^{\prime})\frac{n_{\mathrm{B}}(x^{\prime}) -  n_{\mathrm{B}}(x) }{i\omega+x^{\prime}-x},\label{11}
\end{equation}
where $n_\mathrm{B}(x)=(e^{x/T}-1)^{-1}$ is the Bose-Einstein distribution function. To approximate this integral, we perform Taylor expansion of $n_{\mathrm{B}}(x)$ and $n_{\mathrm{B}}(x^{\prime})$ near the maxima $x=E_{\alpha}, x^{\prime} = E_{\beta}$ of the spectral functions. After that, integration over $x,x^{\prime}$ and analytical continuation  $i\omega \to \omega + i0$ yield the approximate retarded $S$-function
\begin{equation}
    S(E_{\alpha},E_{\beta}) \\ =  \frac{n_{\mathrm{B}}(E_{\alpha}) - n_{\mathrm{B}}(E_{\beta})- i \gamma \left[n^{\prime}_{\mathrm{B}}(E_{\alpha}) + n^{\prime}_{\mathrm{B}}(E_{\beta})\right] }{E_{\alpha} - E_{\beta} -\omega - 2i\gamma},\label{approx}
\end{equation}
which will be used in the following. Numerical verification of this approximation proves its accuracy in the considered parameter ranges for $\alpha \ne \beta$. In contrast, at $\alpha = \beta$ this approximation lacks quantitative accuracy in the Hagen-Rubens regime $\omega \ll \gamma$, although it provides qualitatively correct results and becomes exact in the clean DC limit $\gamma = 0$, $\omega\to0$ (assumed, e.g., in the superfluid drag calculations in Ref.~\cite{Fil2005}).

We will limit ourselves to the case of relatively weak damping $\gamma$ to maintain applicability of the quasiparticle description. Similarly to the Mott-Ioffe-Regel bound \cite{Lee}, validity of quasiparticle description requires the mean free path $l=c_i/\gamma$ of quasiparticles (with  their characteristic velocities $c_i=\sqrt{\mu_{i}/m_{i}}$) being larger than the mean interparticle distance $n_{i}^{1/3}$. Expressing $n_{i}$ through the $T_\mathrm{c}^i$, we obtain restriction for the damping rate $\gamma \ll \sqrt{\mu_{i}T_\mathrm{c}^i}$. Fortunately this condition allows us to consider the system in ballistic regime and neglect the vertex corrections, because the ballistic approximation is appropriate whenever $\bar{p}l>1$  \cite{Kamenev}, where $\bar{p}$ is the characteristic momentum of quasiparticles defined in the next section. Roughly estimating this momentum as $\bar{p}\sim\sqrt{mT}$ (see Appendix~\ref{app:Delta}), we obtain $\bar{p}l\sim \sqrt{\mu_{i}T}/\gamma$. At low enough damping rate assumed above, we obtain $\bar{p}l$ much larger than the ratio $\sqrt{T/T_\mathrm{c}^i}$, which is expected to be of the order of unity at moderate temperatures $T\sim \frac13T_\mathrm{c}^i$ taken in our calculations. Thus our neglect of the vertex corrections is consistent in the assumed range of parameters.

\section{Analytical approximations}\label{sec:Analytic}

\subsection{Contributions of quasiparticle branches}

In this section we develop analytical approximations for frequency-dependent conductivities $\sigma_{ij}^{\mathrm{n}}(\omega), \sigma_{ij}^{\mathrm{s}}(\omega)$, which resemble the familiar Drude and Lorentz models. Our analysis is applicable to 3D atomic mixtures with short-range interactions in the temperature range $\mu_{i}<T\ll T_{\mathrm{c}}$. Inserting the approximate $S$-function (\ref{approx}) into Eq.~(\ref{chiSF}) and performing momentum integration, we obtain
\begin{align}
    \sigma_{ij}(\omega) &\approx \frac{i}\omega\left\{ \frac{\delta_{ij}n_{i}}{m_{i}} - D_{ij}^{0} \right.\nonumber\\ &\left.  -\frac{(-1)^{\delta_{ij}}}{m_{i}m_{j}}\left[ \Lambda^{+}(\omega)+\Lambda^{-}(\omega) \right] \right\} + \frac{i D_{ij}^{0}}{\omega+2i\gamma}.\label{chiSF2}
\end{align}
Here the $D_{ij}^{0}$ terms describe processes where quasiparticles are scattered from one branch into the same branch [density-to-density and spin-to-spin, see Figs.~\ref{fig:Edif}(c-d)], and the corresponding conductivity weights are defined as
\begin{equation}
D^{0}_{ij}=-\sum_{\mathbf{p}} \frac{p^{2}}{2Adm_{i}m_{j}}  \left[P_{i\mathrm{d}}P_{j\mathrm{d}}n^{\prime}_{\mathrm{B}}(E_{\mathrm{d}})+P_{i\mathrm{s}}P_{j\mathrm{s}}n^{\prime}_{\mathrm{B}}(E_{\mathrm{s}})\right].\label{D0_weight}
\end{equation}
The coefficients $P_{i\alpha}$, quantifying contribution of the $i$th component to Bogoliubov excitation branch $\alpha$, are defined by Eq.~(\ref{P_coeff}). The expression (\ref{D0_weight}) is the counterpart of conventional Landau formula for density of the normal component \cite{Landau, Baym1968} generalized for a two-component superfluid system.

Two other terms $\Lambda^\pm(\omega)$ depend on frequency and cannot be calculated analytically, so we derive approximations for them. The function $\Lambda^{+}(\omega)$ is responsible for the processes of  quasiparticle scattering with interconversion from one branch to the distinct one [spin-to-density and vice versa, see Fig.~\ref{fig:Edif}(e)]. The second function $\Lambda^{-}(\omega)$ corresponds to creation or annihilation of two quasiparticles of different branches [Fig.~\ref{fig:Edif}(f)]; note that similar same-branch processes are forbidden at $\mathbf{q}=0$. These functions can be written as momentum integrals
\begin{equation}
\Lambda^{\pm}(\omega) =  \int\limits_{0}^{\infty}dp\left[f_{\pm}(p)R_{\pm}(\omega,p) + f^{*}_{\pm}(p)R^{*}_{\pm}(-\omega,p)\right],\label{app_f}
\end{equation}
where $f_{\pm}(p)$ are defined in Appendix \ref{app:Details} and will be called \emph{envelope functions}, while the \emph{resonant functions} are defined as
\begin{equation}
    R_{\pm}(\omega,p)=\frac{1}{E_{\mathrm{d}}(p) \mp E_{\mathrm{s}}(p)-\omega-2i\gamma}.\label{R_pm}
\end{equation}
The envelope functions $f_{\pm}(p)$ endure power-law increase at low momenta and decrease exponentially at $E_{\mathrm{d,s}}\gtrsim T$ thanks to the Bose-Einstein distribution functions, so they have extrema at some momentum $\bar{p}$ where the quasiparticle energies match the temperature. Therefore it is convenient to define the characteristic momentum $\bar{p}$, whose neighbourhood provides the major contribution to the integral, as solution of equation $E_{\mathrm{d}}(\bar{p})+E_{\mathrm{s}}(\bar{p})=2T$ (see more detailed discussion in Appendix~\ref{app:Delta}). Thus the characteristic sum of quasiparticle energies entering $R_-$ is of the order of $T$. The characteristic difference of energies entering $R_+$ is the important energy parameter
\begin{equation}
\Delta=E_{\mathrm{d}} (\bar{p}) - E_{\mathrm{s}} (\bar{p}),\label{Delta_16}
\end{equation}
which has a meaning of resonance frequency for quasiparticle inter-conversion processes [Fig.~\ref{fig:Edif}(e)] giving rise to the Lorentz-type response at moderately high $\omega$. Depending on relationship between $\omega$ and $\Delta$, we can separate the Drude and Lorentz regimes.

\begin{figure}[t]
\begin{center}
\includegraphics[width=1\columnwidth]{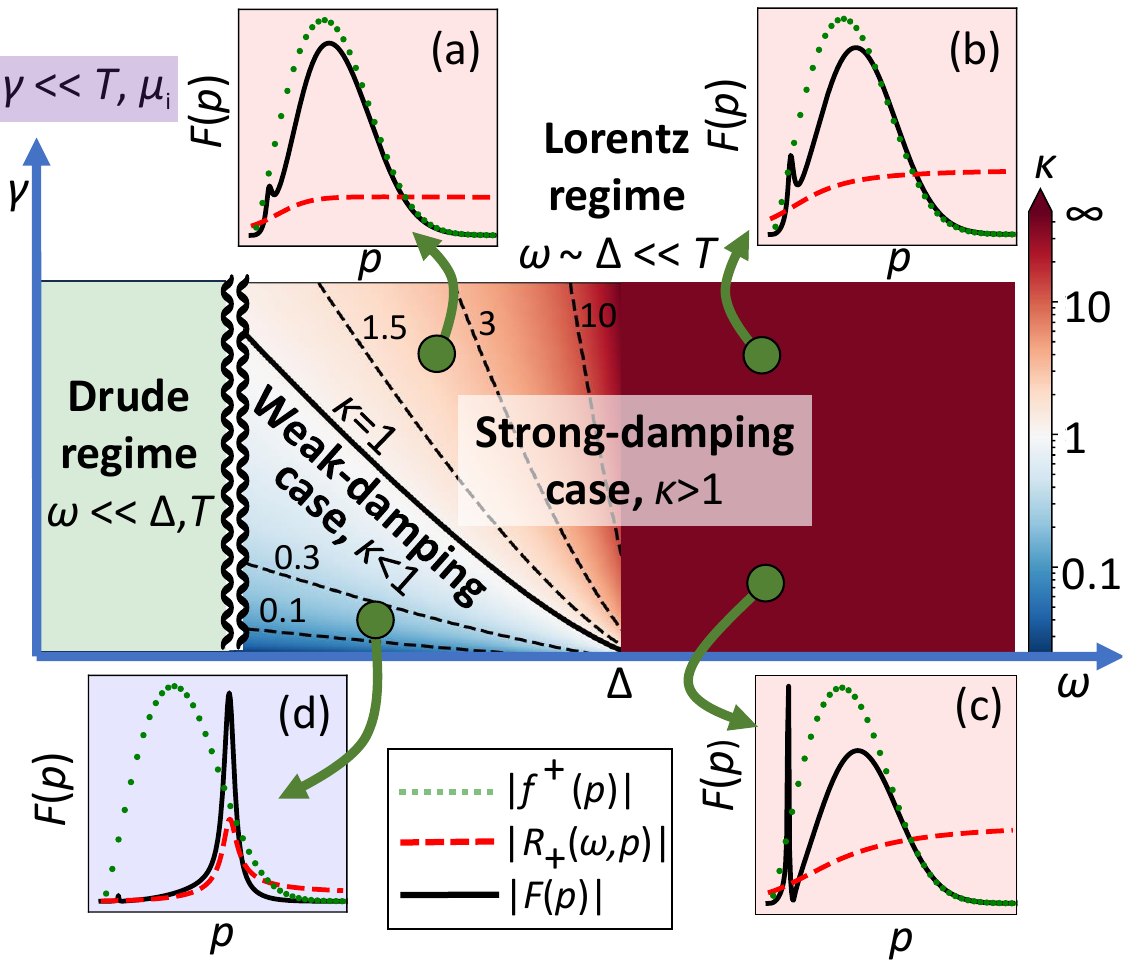}
\end{center}
\caption{Schematic depiction of different regimes on the $\omega$, $\gamma$ plane: the Drude regime at low frequencies and the Lorentz regime at higher frequencies. According to the values of the dimensionless damping parameter $\kappa$ shown by color and contour lines, we separate the Lorentz regime into weak- ($\kappa<1$) and strong-damping ($\kappa>1$) cases whose boundary is shown by thick black line. Insets (a-d) show how $f_{+}(p)$, $R_{+}(p)$, and the total integrand $F(p)=f_{\pm}(p)R_{\pm}(\omega,p) + f^{*}_{\pm}(p)R^{*}_{\pm}(-\omega,p)$ in Eq.~(\ref{app_f}) behave, by absolute value, as functions of $p$.}
\label{fig:Regimes}
\end{figure}

\subsection{Drude regime}

The Drude regime occurs when $\omega$ is far lower than the resonance frequencies $E_{\mathrm{d}}\pm E_{\mathrm{s}}$ in denominators of $R_\pm$. According to the estimates above, it corresponds to the frequency range $\omega\ll\Delta,T$ shown by green shading in Fig.~\ref{fig:Regimes}. In this limit we assume $R_{\pm}(\omega,p)\approx R_{\pm}(0,p)$ in the integrals (\ref{app_f}), so the functions $\Lambda^{\pm}(\omega)$ become almost frequency-independent, and we obtain the simple expression for conductivities in the Drude regime:
\begin{equation}
    \sigma_{ij}(\omega) \approx \frac{i}{\omega}\left\{\frac{\delta_{ij}n_{i}}{m_{i}} - D_{ij}^{0} + D_{ij}^{+}+D_{ij}^{-} \right\}+ \frac{i D_{ij}^{0}}{\omega+2i\gamma}.\label{Sigma_Drud_and_Sp}
\end{equation}
Here 
\begin{align}
    D^{\pm}_{ij}&=-\frac{(-1)^{\delta_{ij}}}{m_{i}m_{j}} \Lambda^{\pm}(0)\nonumber\\
    &=\mp\frac{(-1)^{\delta_{ij}}}{m_{i}m_{j}}  \int\limits_{0}^{\infty} dp \, \mathrm{Re}\,\frac{2f_{\pm}(p)}{E_{\mathrm{d}} \mp 
    E_{\mathrm{s}}-2i\gamma}.\label{26}
\end{align}

Note that the terms $D_{ij}^\pm$, being almost real, contribute mainly to the nondissipative part of the conductivities (\ref{Sigma_Drud_and_Sp}), because low frequencies $\omega$ are far off-resonant from the absorption processes, corresponding to these terms and depicted in Fig.~\ref{fig:Edif}(e-f). Totally, the diamagnetic $n_i/m_i\omega$ and paramagnetic $-D^0_{ij}+D^+_{ij}+D^-_{ij}$ terms in the braces of Eq.~(\ref{Sigma_Drud_and_Sp}) do not cancel each other in the Bose-condensed regime giving rise to the uncompensated $i/\omega$ singularity of both intra- and transconductivities in the DC limit $\omega\rightarrow0$. Amplitudes of such singularities are characterized by superfluid weights \cite{Scalapino} (which are proportional to the density of superfluid component or to the inverse square of the London penetration depth in the case of superconductors) $D^{\mathrm{s}}_{ij}=\pi(\delta_{ij}n_i/m_i-D_{ij}^0+D_{ij}^++D_{ij}^-)$. On the other hand, the Drude weights, characterizing an integral low-frequency ability of a conductor to maintain the dissipative conductivity and defined as $2\int_0^\infty d\omega\mathrm{Re}\,\sigma_{ij}(\omega)$, in our case are equal to $D^{\mathrm{n}}_{ij}=\pi D_{ij}^0$.

\begin{figure}[t]
\begin{center}
\includegraphics[width=1\columnwidth]{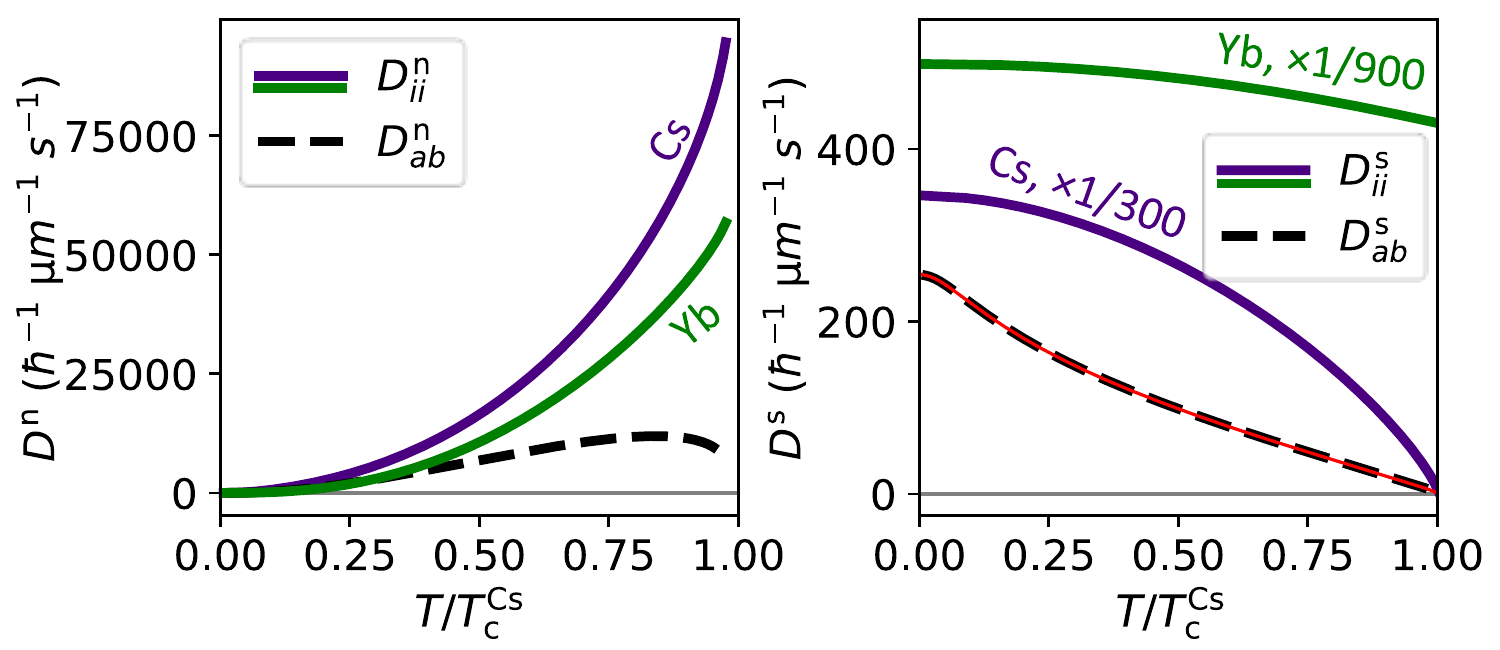}
\end{center}
\caption{Temperature dependencies of the Drude (left panel) and superfluid (right panel) weights for Yb-Cs mixture at $\gamma/2\pi=1\,\mbox{Hz}$. Solid lines correspond to the intracomponent, and black dashed line to the intercomponent conductivities. Thin red line shows the approximation $D_{ab}^\mathrm{s}\approx\pi\rho_{\mathrm{dr}}/m_{a}m_{b}$, where $\rho_\mathrm{dr}$ is the drag density calculated at $\gamma=0$ \cite{Fil2005}. Calculation parameters are listed in Table~\ref{table:1}.}
\label{fig:Weights}
\end{figure}

The example of temperature dependencies of Drude and superfluid weights is shown in Fig. \ref{fig:Weights} for the mass-imbalanced Yb-Cs mixture. As expected, Drude weights vanish at $T=0$, when the mixture is fully in superfluid state, and superfluid weights involving Cs subsystem vanish at $T=T_\mathrm{c}^\mathrm{Cs}$ when it becomes normal. We also notice that our results at low $\gamma$ are in agreement with the theory of DC superfluid drag developed for clean systems (red thin line) by Fil and Shevchenko \cite{Fil2005}. The recession of the intercomponent Drude weight $D^n_{ab}$ down to zero near $T=T_\mathrm{c}^\mathrm{Cs}$ is the artefact of our one-loop approximation, which neglects more complicated diagrams contributing to drag in the normal state \cite{Kamenev}. However, they can be neglected at low enough temperatures $T\lesssim T_{\mathrm{c}}^i$ \cite{Aminov_Quantum}.

\subsection{Lorentz regime}

The Lorentz regime occurs when $\omega$ is close to the resonant energy $\Delta$ of the spin-to-density quasiparticle conversion. In this regime only $R_{+}$ demonstrates a resonance behavior and becomes dominant, and the other function $R_{-}$ can be neglected, because $E_{\mathrm{d}} + E_{\mathrm{s}}\gg E_{\mathrm{d}} - E_{\mathrm{s}}$ at typical momentum $\bar{p}$. Also we may notice that the $f_{+}(p)R_{+}(\omega,p)$ term in the integral (\ref{app_f}) is dominant over the off-resonant term $f_{+}^{*}(p)R^{*}_{+}(-\omega,p)$.

We subdivide the Lorentz regime into weak- and strong-damping cases, depending on the dimensionless parameter $\kappa=\Delta p/\bar{p}$, defined as the related to $\bar{p}$ momentum width $\Delta p \approx 4\gamma/[E^{\prime}_{\mathrm{d}}(p_{+})-E^{\prime}_{\mathrm{s}}(p_{+})]$ of the resonant function $R_{+}(\omega,p)$ around its maximum at $p=p_{+}$, which characterizes both the maximum and the typical width of the envelope function $f_{+}(p)$. As shown in Fig. \ref{fig:Regimes}, the weak-damping case $\kappa\ll1$ [Fig.~\ref{fig:Regimes}(d)] means that $R_{+}(\omega,p)$ is very narrow along the momentum axis, in comparison with $f_{+}(p)$. In the strong-damping case $\kappa\gg1$, as depicted in Figs.~\ref{fig:Regimes}(a-c), the situation is opposite. We assign the frequency region $\omega>\mu_a+\mu_b$, where $R_+$ is never resonant and monotonously increases (so $\Delta p$ is undefined), to the strong-damping case as well, setting formally $\kappa=\infty$ in this region.

\subsubsection{Weak-damping case}\label{Sec_weak_damping}

In the weak-damping case, when $R_{\pm}(\omega,p)$ is very narrow, we can bring $\gamma$ in its denominator to zero and find $\Lambda_+(\omega)$ analytically by integration of the resulting Dirac delta function in Eq.~(\ref{app_f}) to obtain
\begin{align}
\mathrm{Re} \, \Lambda^{+} (\omega) &\approx 2 \int\limits_{0}^{\infty} dp\,f_{+}(p)\frac{E_{\mathrm{d}}(p)-E_{\mathrm{s}}(p)}{[E_{\mathrm{d}}(p)-E_{\mathrm{s}}(p)]^{2}-\omega^{2}},\label{clean1}\\
\mathrm{Im} \, \Lambda^{+} (\omega) &\approx-\frac{\pi f_{+}(p_{+})}{E^{\prime}_{\mathrm{d}}(p_{+})-E^{\prime}_{\mathrm{s}}(p_{+})},\label{clean2}
\end{align}
where $p_{+}$ is solution of equation $E_{\mathrm{d}} - E_{\mathrm{s}}=\omega$ dependent on $\omega$, i.e. the momentum where $|R_+(p,\omega)|$ attains sharp maximum; we also assume $\gamma=0$ in the expression (\ref{20}) for $f_+(p)$. The result (\ref{clean2}) for the function $\mathrm{Im} \, \Lambda^{+} (\omega)$, related to the dissipation spectrum $\mathrm{Re}\,\sigma_{ij}(\omega)$, may be interpreted as the sharp resonant function $R_{+}(\omega,p)$ scanning the broad envelope function $f_{+}(p)$ when $\omega$ is changed. The derivative $dp_+/d\omega = [E^{\prime}_{\mathrm{d}}(p_{+})-E^{\prime}_{\mathrm{s}}(p_{+})]^{-1}$ determines the scanning speed along the $p$ axis and hence magnitude of $\mathrm{Im}\,\Lambda^+(\omega)$. The resonance maximum of $\mathrm{Re}\,\sigma_{ij}(\omega)$ is located at $p_{+}=\bar{p}$, where the maxima of two functions $R_{+}(\omega,p)$ and $f_{+}(p)$ coincide. Shape of this resonance depends on the ratio of atomic masses $m_a$, $m_b$ of two components. As discussed in more detail in Appendix~\ref{app:Delta}, we outline two cases: when atomic masses are close to each other (and, in particular, equal in the case of spin mixtures), and when they are distant. These cases are distinguished by how the energy difference $E_{\mathrm{d}} - E_{\mathrm{s}}$ depends on $p$ in the relevant range of momenta $p\sim\bar{p}$. In case of close masses this difference is almost constant [Fig.~\ref{fig:Edif}(a)], $E_{\mathrm{d}} - E_{\mathrm{s}}\approx\mu_{a}+\mu_{b}$, thus the scanning speed $dp_{+}/d\omega$ is high, and the resonance is sharp (see Fig.~\ref{fig:tr} in the next section). In the case of distant masses the energy difference retains an essential momentum dependence, $E_{\mathrm{d}} - E_{\mathrm{s}} \propto p^{2}$ [Fig. \ref{fig:Edif}(b)], so $dp_{+}/d\omega$ is low, and the resonance becomes strongly smeared or vanishes completely (see results for Yb-Cs mixture in Fig.~\ref{fig:all} below).

\subsubsection{Strong-damping case}

In the strong-damping case, the resonant function $R_{+}(\omega,p)$ is much wider than the envelope $f_{+}(p)$ [$\kappa \gg 1$, see Figs.~\ref{fig:Regimes}(a-c)], so we approximate $R_{+}(\omega,p)$ by $R_{+}(\omega,\bar{p})$ to obtain
\begin{align}
    \Lambda^{+}(\omega)\approx \left( \frac{1}{\Delta-\omega-2i\gamma} + \frac{1}{\Delta+\omega+2i\gamma} \right)\int\limits_{0}^{\infty}dp\,f_{+}(p).\label{Lambda}
\end{align}
Here we neglected the $\gamma n^{\prime}_{\mathrm{B}}$ terms in the $f_{+}(p)$ function (\ref{20}), since their order is $\gamma/T\ll1$ (however in the Drude regime these terms should be retained, see Appendix~\ref{app:Details}).

The conductivity in this case is predominantly determined by the first term in the parentheses of Eq.~(\ref{Lambda}) which is resonant near $\omega=\Delta$ with the width $2\gamma$. The role of the second non-resonant term is to red-shift and broaden this resonance.

\section{Numerical calculations}\label{sec:NumCalc}

In this section we present numerical results for the conductivities for various systems and compare them with analytical approximations. The numerically calculated conductivities are found using Eqs.~(\ref{basis}), (\ref{chiSF}) with the approximation (\ref{approx}) for the $S$-function, which proves to be quite accurate in the considered range of parameters. The analytical approximations are given by Eq.~(\ref{Sigma_Drud_and_Sp}) in the Drude regime and Eq.~(\ref{chiSF2}) in the Lorentz regime, with $\Lambda^-=0$ and $\Lambda^+$ given by Eqs.~(\ref{clean1})--(\ref{clean2}) in the weak-damping case ($\kappa<1$) and by Eq.~(\ref{Lambda}) in the strong-damping case ($\kappa>1$). With the considered atomic gases, the weak-damping case is realized at typical damping rates $\gamma/2\pi<10\,\mbox{Hz}$, and the strong-damping case arises at $\gamma/2\pi>100\,\mbox{Hz}$. For each atomic mixture, we take the temperature $T=\frac13\mathrm{min}[T_\mathrm{c}^a,T_\mathrm{c}^b]$, which is low enough for the Bogoliubov approximation to be applicable yet still experimentally feasible.

\begin{figure}[t]
\begin{center}
\includegraphics[width=1\columnwidth]{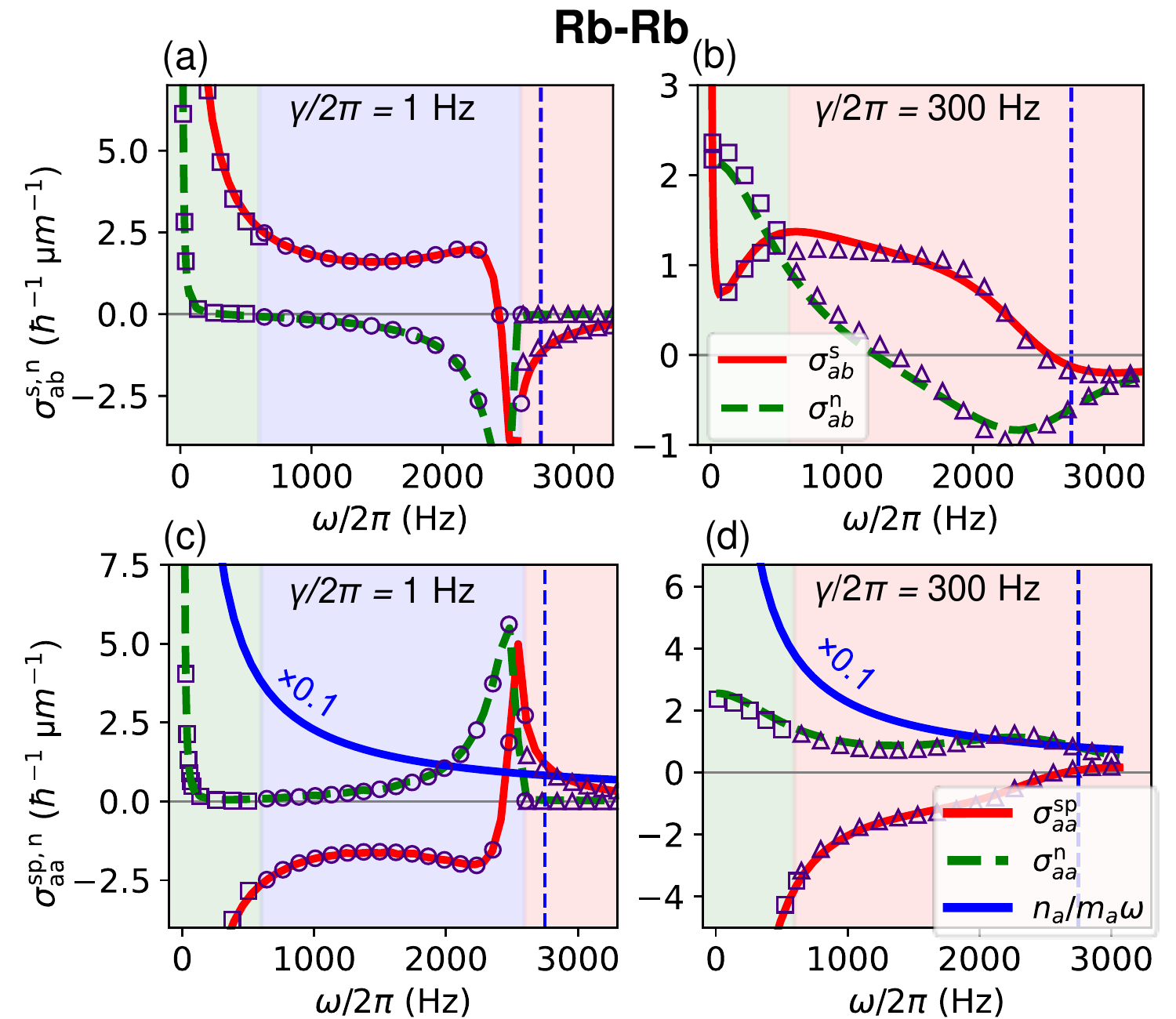}
\end{center}
\caption{Trans- (a,b) and intercondictivity (c,d) for Rb-Rb spinor mixture at weak (left panels) and strong (right panels) damping $\gamma$. Solid and dashed lines show numerical calculations, and symbols show analytical approximations for appropriate regimes depicted by the same color shadings as in Fig.~\ref{fig:Regimes}: Drude regime (squares, green), weak-damping Lorentz regime (circles, blue), and strong-damping Lorentz regime (triangles, red). Vertical dashed lines indicate the frequency $\omega = \mu_{a}+\mu_{b}$, which is close to the resonance frequency $\Delta$. In the intracomponent channel (c,d) the superfluid conductivity $\sigma^\mathrm{s}_{aa}$ is separated into dia- ($n_a/m_a\omega$) and paramagnetic ($\sigma^\mathrm{sp}_{aa}$) parts. Calculation parameters are listed in Table~\ref{table:1}.}  \label{fig:tr}
\end{figure}

In Fig.~\ref{fig:tr} we show the trans- and intraconductivities for the symmetric Rb-Rb mixture at weak [Fig.~\ref{fig:tr}(a,c)] and strong [Fig.~\ref{fig:tr}(b,d)] damping. At low frequencies $\omega \ll \gamma$, the superfluid conductivities $\sigma^\mathrm{s}_{ij}$ are positive and diverge as $1/\omega$ (in the intracomponent channel $i=j$ the positive diamagnetic part $n_a/m_a\omega$ dominates the negative paramagnetic part $\sigma^\mathrm{sp}_{aa}$). It is a signature of nonzero and positive drag (\ref{rho_dr}) and superfluid (\ref{rho_s}) densities. The normal conductivities $\sigma^{\mathrm{n}}_{ij}$ tend to constants in DC limit in conformity with traditional normal Coulomb drag effect and Drude theory of conductivity, although in the weak-damping case [Fig.~\ref{fig:tr}(a,c)] their levelling off at $\omega\rightarrow0$ is not visible at the chosen scale, because the Drude peaks are much higher than the Lorentz-regime features which we are concentrating on. At higher frequencies near $\omega = \Delta$ (which is $\Delta\approx\mu_{a}+\mu_{b}$ when $m_{a}=m_{b}$), absolute value of the normal conductivity $\sigma^{\mathrm{n}}_{ij}$ exhibits absorption peak, while the superfluid transconductivity $\sigma^{\mathrm{s}}_{ab}$ and paramagnetic part $\sigma^{\mathrm{sp}}_{aa}$ in the intracomponent channel change sign. Such features resemble resonant behaviour of the Lorentz model [Fig.~\ref{fig:tr}(c,d)], although in the intracomponent channel this resonant-like behavior of the paramagnetic superfluid conductivity (\ref{sigma_s}) is masked by the large and monotonously decreasing diamagnetic term. 

At frequencies near the resonance $\Delta$, the conductivities are related to each other via $m_{i}^{2}\sigma^{\mathrm{sp,n}}_{ii}(\omega)\approx -m_{a}m_{b}\sigma^{\mathrm{s,n}}_{ab}(\omega)$. This is evident in Fig.~\ref{fig:tr} where $\sigma_{aa}^\mathrm{sp}(\omega)\approx-\sigma_{ab}^\mathrm{s}(\omega)$ and $\sigma_{aa}^\mathrm{n}(\omega)\approx-\sigma_{ab}^\mathrm{n}(\omega)$ near the resonance, since $m_a=m_b$. This feature follows from Eq.~(\ref{chiSF2}) if we omit the terms $iD^{0}_{ij}/(\omega+2i\gamma)$ and $-iD_{ij}^0/\omega$, whose contribution is diminished at large frequencies.

\begin{figure}[t]
\begin{center}
\includegraphics[width=1\columnwidth]{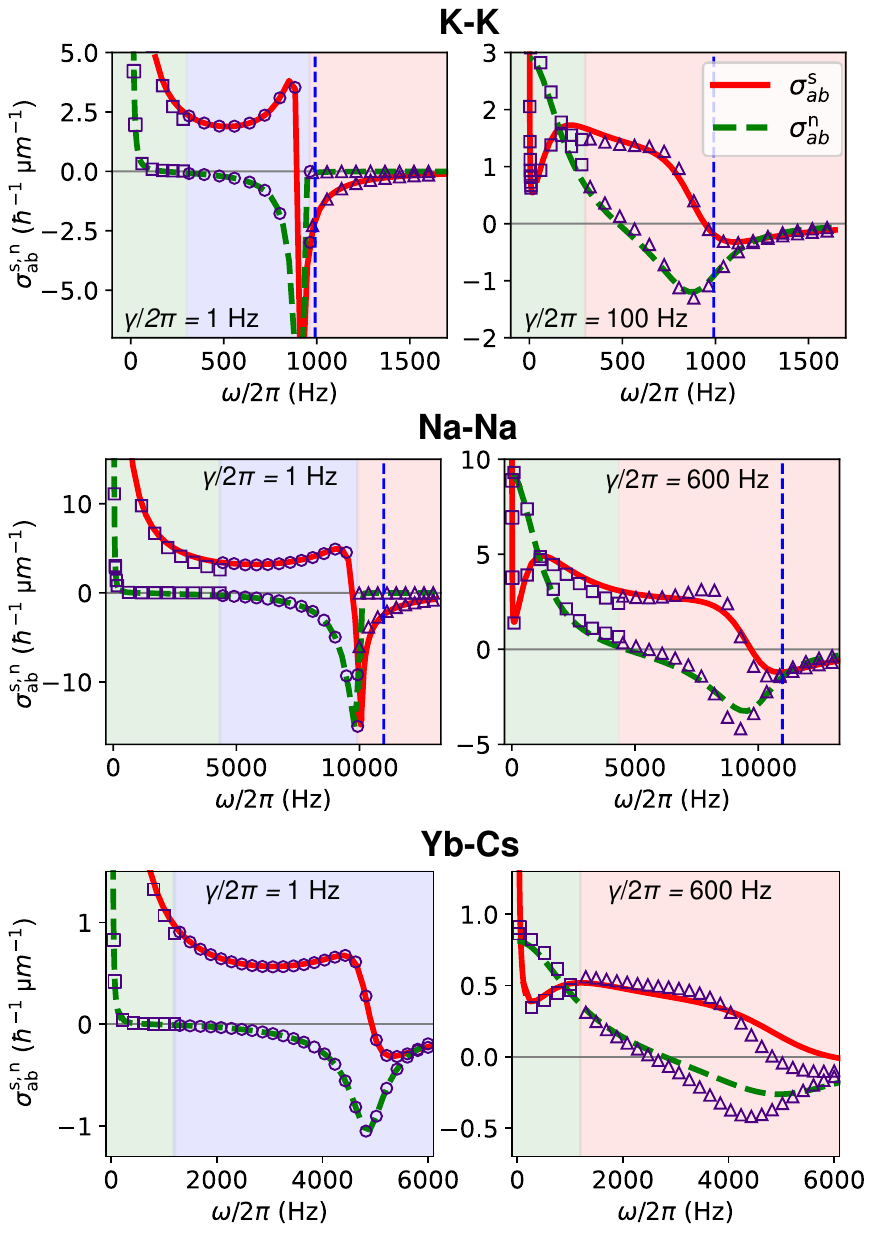}
\end{center}
\caption{Transconductivities of K-K, Na-Na, and Yb-Cs mixtures, each calculated at two values of $\gamma$, where the weak- or strong-damping cases develop in the Lorentz regime. Designations of curves and symbols are the same as in Fig.~\ref{fig:tr}. Calculation parameters are listed in Table~\ref{table:1}.}
\label{fig:all}
\end{figure}

To get more insight into behavior of transconductivities, in Fig.~\ref{fig:all} we show them for equal-mass K-K, Na-Na mixtures and for mass-imbalanced mixture Yb-Cs. It can be seen that the transconductivity of the non-symmetric spin mixture K-K with relatively low resonance energy $\Delta$ exhibits the same features as for Rb-Rb mixture: Drude peak at low frequencies and resonance at $\omega\approx\Delta$. In the case of symmetric spin mixture Na-Na, the resonance frequency $\Delta$ is higher (more than 10 kHz) and presumably out of reach of present experiments capabilities. For the mass-imbalanced mixture Yb-Cs, the resonance frequency $\Delta$ turns out to be much lower than $\mu_{a}+\mu_{b}$, but the resonance itself is degraded in both weak- and strong-damping cases by the reasons discussed in Sec.~\ref{Sec_weak_damping}.

\begin{figure}[t]
\begin{center}
\includegraphics[width=1\columnwidth]{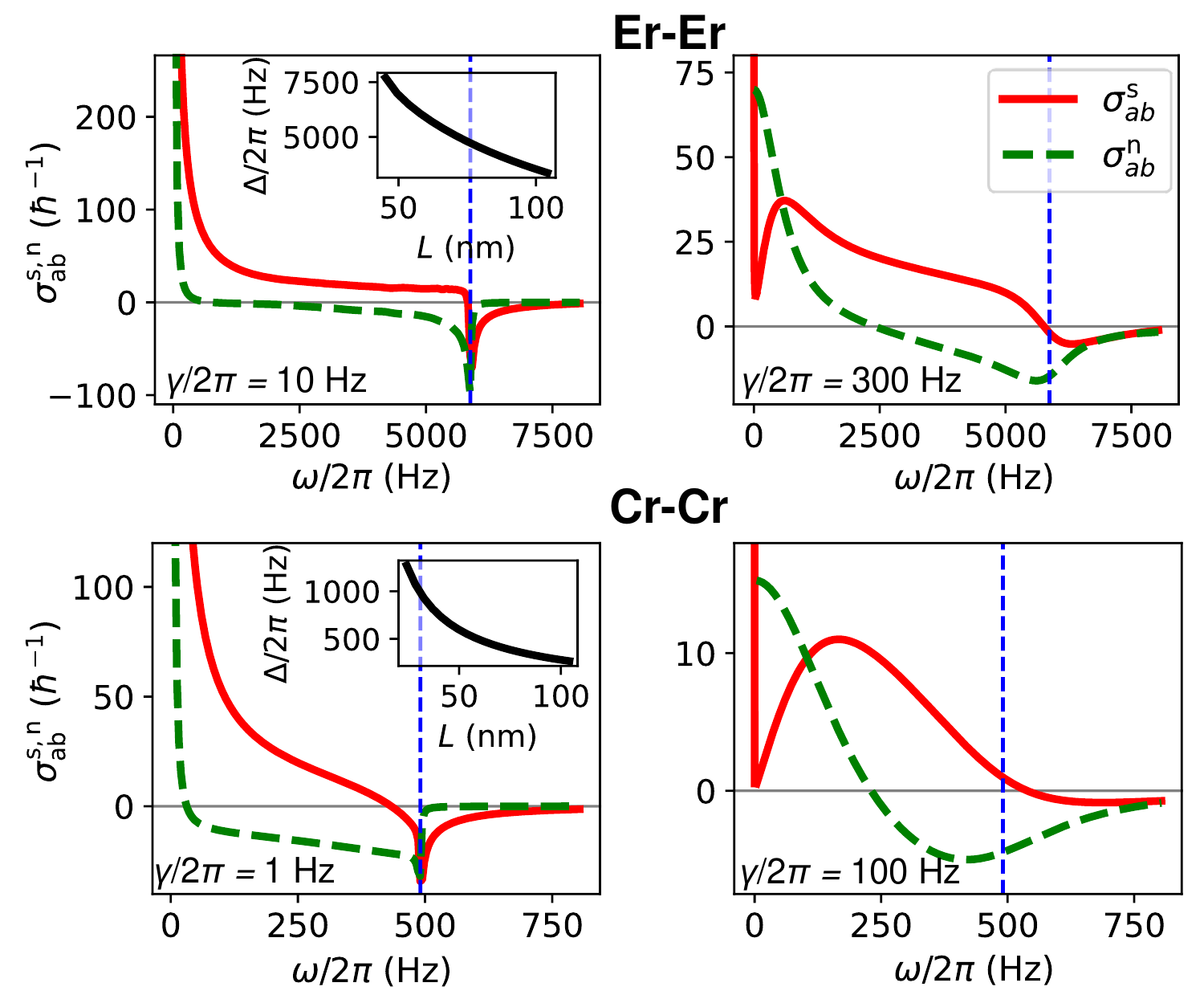}
\end{center}
\caption{Transconductivity between spatially separated clouds of dipolar atoms Er-Er and Cr-Cr in the systems at weak (left panels) and strong (right panels) damping with $L=60\,\mbox{nm}$. Vertical dashed lines indicate the resonance frequency $\omega = \Delta$. Insets show dependence of $\Delta$ on interlayer distance $L$. Thicknesses of both clouds are $w_{z}=20\,\mbox{nm}$, and other calculation parameters are listed in Table~\ref{table:1}.}  \label{fig:tr-dd}
\end{figure}

In Fig.~\ref{fig:tr-dd} we present numerically calculated transconductivities for the pairs of dipolar atomic gases Er-Er and Cr-Cr arranged into quasi-2D two-layered systems [see Fig.~\ref{fig:Scheme}(b)]. The analytical approximations are not applied in this case due to different form of intercomponent dipole-dipole interaction which retains essential momentum dependence, as discussed in Appendix~\ref{app:d-dint}. In contrast to 3D mixtures with short-range interactions, here we can tune the resonance frequency $\Delta$ by varying the interlayer distance $L$. This frequency, found as the maximum of the quasiparticle energy difference $\Delta=\max[E_{\mathrm{d}}(p) - E_{\mathrm{s}}(p)]$, approximately follows the $\Delta\propto L^{-1}$ trend, as shown in the insets in Fig.~\ref{fig:tr-dd}. The intraconductivities in this case are not shown, since the relation between total $n_i$ and condensate $n^{0}_{i}$ densities, needed to describe partial compensation of the diamagnetic term with quantitative accuracy, is not well-defined in 2D systems in the framework of Bogoliubov theory, and more complicated approaches, such as quasicondensate analysis \cite{Prokofiev}, should be applied, which is beyond the scope of our paper.

\section{Discussion}\label{sec:Discussion}

In this paper, we studied intra- and transconductivities $\sigma_{ij}(\omega)$ of a homogeneous two-component superfluid Bose-condensed systems at nonzero frequencies $\omega$. We calculated the conductivities in one-loop approximation using the Bogoliubov theory of a two-component BEC at finite temperature and with taking into account the phenomenological damping $\gamma$ of spin and density quasiparticle modes. Two possible setups of the two-component atomic system are considered: 3D spinor atomic mixtures (Rb-Rb, K-K, Na-Na, Yb-Cs) and spatially separated two-layered systems with magnetic dipole-dipole interactions (Er-Er and Cr-Cr). 

We separate each conductivity $\sigma_{ij}(\omega)$ into the real part $\sigma_{ij}^\mathrm{n}(\omega)$, which is responsible for dissipative response (current in phase with a driving force), and imaginary part $\sigma_{ij}^\mathrm{s}(\omega)$, which corresponds to non-dissipative response with the $\pi/2$ phase delay. Our analysis shows that, at frequencies much lower than the characteristic energy gap $\Delta$ between spin and density quasiparticle modes, the conductivities are well described by the two-fluid Drude model \cite{Brorson_1996,Yang_2018} where $\sigma_{ij}^\mathrm{n}(\omega)$ exhibits the Drude peak $\propto[\omega^2+4\gamma^2]^{-1}$ like the normal metallic conductivity (in the intracomponent channel $i=j$) or normal drag effect (in the intercomponent channel $i\neq j$), while $\sigma_{ij}^\mathrm{s}(\omega)$ demonstrates the $1/\omega$ singularity indicating superfluidity (at $i=j$) or superfluid drag effect (at $i\neq j$). Thus our theory describes dissipative conductivity, superfluidity, as well as normal and superfluid drag effects on equal footing.

At higher frequencies near $\omega \sim \Delta$ the dissipative part of conductivity $\sigma_{ij}^{\mathrm{n}}(\omega)$ exhibits peak, while non-dissipative part $\sigma_{ij}^{\mathrm{sp}}(\omega)$ changes sign, which is qualitatively similar to the Lorentz model of resonant response. However in our case the resonance shape is asymmetric and can essentially differ depending on the damping rate $\gamma$ and whether the atomic masses in a mixture are close to each other or distant enough. In a symmetric mixture with equal masses, $\Delta$ is close to the sum $\mu_a+\mu_b$ of atomic chemical potentials, and the general case is considered in Appendix~\ref{app:Delta}. For two-layered quasi-2D system of dipolar atoms, $\Delta$ can be tuned by varying the interlayer separation $L$. For 3D mixtures [Fig.\ref{fig:Scheme}(a)], we derive the analytical formulas which approximate the conductivities in both Drude and Lorentz regimes rather accurately.

All considered examples of 3D and quasi-2D atomic two-component systems demonstrate similar features of their conductivity spectra: $1/\omega$ singularities of $\mathrm{Im}\,\sigma_{ij}(\omega)$ signaling superfluidity and superfluid drag, Drude-like peaks of finite heights in $\mathrm{Re}\,\sigma_{ij}(\omega)$ at low frequencies responsible for normal dissipative conductivity and normal drag on top of superfluidity, and deformed Lorentz-like resonances at higher frequencies, which significantly broaden at high quasiparticle damping rate or at large atomic mass imbalance. These features are well described both qualitatively and quantitatively by the obtained analytical approximations for $\sigma_{ij}(\omega)$.

In our calculations we assumed the momentum- and energy-independent damping rate $\gamma$, which allowed to obtain analytically tractable results. Theoretical analysis show that Beliaev and Landau damping of Bogoliubov quasiparticles, which is caused by scattering on other thermally excited quasiparticles, increases with momentum \cite{Pitaevskii_1997,Chung_2009}. Damping due to scattering on external disorder generally increases with momentum as well \cite{Giorgini_1994,Lopatin_2002,Gaul_2011,Muller_2012}, however at strong enough disorder the relaxation kernel, which plays the role of damping rate in the Drude-like formula for conductivity, acquires the $1/\omega$ singularity signifying transition to Anderson insulating phase \cite{Gold}. Our calculations show that dominating contribution to the conductivities is provided by quasiparticles with characteristic momenta $p\sim\bar{p}$, so we can approximate $\gamma$ by the damping rate of quasiparticles in vicinity of this momentum.

The drag effects predicted in our paper can be observed in experiments with two-component or two-layered atomic BECs by detecting currents arising in response to an alternating force, which selectively drives one of the components (or drives them in opposite directions). The currents can be determined by measuring atomic velocities via atomic cloud imaging after trap release or by time-of-flight measurements. The driving force can be imposed by magnetic field gradients \cite{Merloti_2013}, optical lattices \cite{Gauthier_2021}, magnetic trap shaking \cite{Llorente_Garc_a_2013}, or sudden displacement of optical trap \cite{DuLi_2023}. Such methods can provide oscillation frequencies up to several kHz, and achievable frequency ranges are often dictated by properties of the atoms themselves \cite{Sekino_2022}. 

Let us estimate a magnetic field gradient required to induce strong enough oscillations, which could be observed by standard atomic cloud imaging. Consider the Yb-Cs atomic mixture \cite{Wilson_2021}, where $^{174}$Yb lacks magnetic moment, so its Lande factor is zero $(g_F=0)$, and thus only $^{133}$Cs is affected by magnetic field ($g_F=-0.25$ \cite{Cesium}). In a homogeneous system the field gradient, required to induce oscillations of the $^{133}$Cs atomic cloud with the amplitude $x_{\mathrm{Cs}}$ and frequency $\omega$, is $|\nabla B| = m_{\mathrm{Cs}}\omega^{2}x_{\mathrm{Cs}}/m_Fg_F\mu_{\mathrm{B}}$, where $m_{\mathrm{Cs}}$ is the mass of $^{133}$Cs atom, $m_F$ is the magnetic sublevel of hyperfine state with angular momentum $F$. Assuming the detectable amplitude $x_{\mathrm{Cs}}\sim 10\,\mu\mbox{m}$ and using parameters from \cite{Wilson_2021}, we obtain $|\nabla B| (\mathrm{G/cm})\approx [(\omega/2\pi)(\mathrm{Hz})]^{2}\times10^{-4}$. The gradients up to 3000 G/cm used in experiments \cite{Vuletic_1996} are sufficient to create oscillations in both Drude ($\omega/2\pi \sim 100\,\mbox{Hz}$, $|\nabla B|\sim1\,\mbox{G/cm}$) and Lorentz ($\omega/2\pi \sim 5000\,\mbox{Hz}$, $|\nabla B|\sim2500\,\mbox{G/cm}$) regimes. The presence of harmonic trap alters relationship between $x_\mathrm{Cs}$ and $\nabla B$, and we may hope to use the mechanical resonance effects to enhance the oscillation amplitude even more.

For reliable detection of the drag effects, we need to achieve large enough amplitude $x_\mathrm{Yb}$ of oscillating motion of the $^{174}$Yb atomic cloud in response to the magnetic field gradient force applied to $^{133}$Cs atoms. The ratio of oscillation amplitudes can be estimated as  $x_{\mathrm{Yb}}/x_{\mathrm{Cs}}=j_{\mathrm{Yb}}n_{\mathrm{Cs}}/j_{\mathrm{Cs}}n_{\mathrm{Yb}}=(n_{\mathrm{Cs}}/n_{\mathrm{Yb}})\times|\sigma_{ab}(\omega)/\sigma_{aa}(\omega)|$. At plausibly low damping rate $\gamma/2\pi=1$ Hz, we obtain $x_{\mathrm{Yb}}/x_{\mathrm{Cs}}\sim 0.01$ at $\omega/2\pi=100$ Hz and $x_{\mathrm{Yb}}/x_{\mathrm{Cs}}\sim 0.05$ at $\omega/2\pi=5000$ Hz. Such ratios are not restrictingly small, so we may hope to detect oscillations of the passive $^{174}\mathrm{Yb}$ component at high enough oscillating force and large enough oscillation amplitudes $x_{\mathrm{Cs}}$ of the active component. For Rb-Rb mixture (see Fig. \ref{fig:tr}) this ratio is generally larger: $x_a/x_b\sim 0.06$ at low frequencies and $x_a/x_b\sim 0.4$ at the Lorentz-like resonance.

Plane-parallel systems of magnetic dipole atoms \cite{Politi_2022,DuLi_2023} possess several additional controllable parameters: thickness of the clouds $w_{z}$, intercloud separation $L$, and dipole moments orientation. The theory presented in our paper allows us to calculate the transconductivity between dipolar atomic clouds in the setup of Ref.~\cite{DuLi_2023}. However, direct comparison of our calculations with results of this experiment is hindered because atomic gases in Ref.~\cite{DuLi_2023} were not Bose-condensed, and harmonic traps used to hold them made the atomic clouds inhomogeneous and prone to mean-field repulsion not described by our theory. It is of interest to extend our approach to take into account the normal-state drag diagrams \cite{Kamenev,Narozhny2016,Aminov_Quantum} which would allow to describe the AC drag in wide temperature range both below and above $T_\mathrm{c}$.

An alternative way to infer information about trans- and intraconductivities can rely on measuring temperature changes after several oscillations \cite{Llorente_Garc_a_2013}. Mutual entrainment of two components can also affect dispersions and damping rates of first and second sounds in the two-component BECs, which can be detected in sound velocity measurements \cite{Hilker_2022}. Our approach of conductivity calculations is aimed on homogeneous systems corresponding to flat traps \cite{Hilker_2022,Yan_2022,Navon}. In harmonic traps the resonance in center-of-mass motion of atomic clouds alters the behaviour of conductivity \cite{Llorente_Garc_a_2013}, and the mean-field repulsion effects mimicking intrinsic interlayer conductivity can appear \cite{Huang,Demin,Matveeva,Wilson_2021}, so the problem of mutual entrainment becomes more complicated.

To conclude, the theory of conductivities of Bose-condensed two-component systems developed in this paper unifies calculations of the normal drag effect, Andreev-Bashkin effect, as well as intracomponent DC conductivity and superfluid density. Investigation of frequency dependencies of the conductivity tensor allows us to study interplay of dissipative and nondissipative current responses. Our approach can be generalized for spin conductivity calculations \cite{Sekino_2022,Sekino_2023} and for coupled 1D atomic gases \cite{Mistakidis}. Besides, similar AC entrainment effects, both dissipative and nondissipative, can be expected in Fermi-atom and condensed-matter superconducting systems.

\section*{Acknowledgments}
The work on analytical calculation and approximation of conductivities was done as a part of research Project No. FFUU-2024-0003 of the Institute for Spectroscopy of the Russian Academy of Sciences. The work on numerical calculations was supported by the Program of Basic Research of the Higher School of Economics.

\appendix
\section{Green functions}\label{app:BT}

The Hamiltonian of homogeneous two-component atomic system is
\begin{align}
    H&=\sum_{i\mathbf{p}}  \epsilon_{i\mathbf{p}} a_{i\mathbf{p}}^{\dagger} a_{i\mathbf{p}}\nonumber\\
    &+\frac1{2A}\sum_{ij\mathbf{p}\mathbf{p}^{\prime}\mathbf{q}}V_{ij}(\mathbf{q})a_{i,\mathbf{p}+\mathbf{q}}^{\dagger} a_{j,\mathbf{p}^{\prime}-\mathbf{q}}^{\dagger} a_{j\mathbf{p}^{\prime}}a_{i\mathbf{p}},
\end{align}
where $a_{i\mathbf{p}}$ is the destruction operator of the atomic particle of the component $i=a,b$ with momentum $\mathbf{p}$, $\epsilon_{i\mathbf{p}}=p^2/2m_i$ is atomic dispersion, and $V_{ij}(\mathbf{q})$ is the Fourier transform of the interaction between particles $i$ and $j$. For 3D atomic mixtures we approximate the interactions by momentum-independent constants $V_{ij}(\mathbf{q})\approx g_{ij} = 2\pi a^\mathrm{s}_{ij} (1/m_{i} + 1/m_{j})$ related to the s-wave scattering lengths given in Table~\ref{table:1}. For magnetic dipolar atoms we take into account additional long-range interactions as shown in Appendix~\ref{app:d-dint} below. Replacing each zero-momentum operator $a_{i,\mathbf{p}=0}$ by square root of the number of condensate particles $(An_{i}^{0})^{1/2}$, we obtain the mean-field Bogoliubov Hamiltonian, which can be further diagonalized by the transformation
\begin{equation}
    a_{i\mathbf{p}}=u^+_{i\mathrm{d}}B_{\mathrm{d}\mathbf{p}}+u^{-}_{i\mathrm{d}}B^{\dag}_{\mathrm{d},-\mathbf{p}}+u^{+}_{i\mathrm{s}}B_{\mathrm{s}\mathbf{p}}+u^{-}_{i\mathrm{s}}B^{\dag}_{\mathrm{s},-\mathbf{p}}\label{Transformation}
\end{equation}
into usual form $\sum_{\mathbf{p}}(E_{\mathrm{d}}B^{\dag}_{\mathrm{d}\mathbf{p}}B_{\mathrm{d}\mathbf{p}}+E_{\mathrm{s}}B^{\dag}_{\mathrm{s}\mathbf{p}}B_{\mathrm{s}\mathbf{p}})$, where $B_{\mathrm{d}\mathbf{p}}$, $B_{\mathrm{s}\mathbf{p}}$ are destruction operators of density and spin quasiparticles. Their energies read 
\begin{equation}
    E_{\mathrm{\mathrm{d},\mathrm{s}}}^2=\frac{E_a^2+E_b^2}{2} \pm \sqrt{\left(\frac{E_a^2-E_b^2}{2}\right)^2+4 \epsilon_a \epsilon_b n_{a}^{0}n_{b}^{0}\left|V_{ab}\right|^{2}},\label{E_dispersion}
\end{equation}
and $E_i$ is the energy of Bogoliubov excitation of isolated $i$th component: $E_i=\sqrt{\epsilon_{i}(\epsilon_{i}+2n_{i}^{0}V_{ii})}$. Here the chemical potentials $\mu_i=n_i^0V_{ii}$ of both components cancel the Hartree mean-field self-energies making both $E_{a,b}$ and $E_\mathrm{d,s}$ gapless. The Bogoliubov transformation coefficients are 
\begin{equation}
u_{a \alpha}^\zeta=\frac{\epsilon_a+\zeta E_\alpha}{2 \sqrt{\epsilon_a E_\alpha}} \sqrt{P_{a \alpha}}, \quad u_{b \alpha}^\zeta=\pm\frac{\epsilon_b+\zeta E_\alpha}{2 \sqrt{\epsilon_b E_\alpha}} \sqrt{P_{b \alpha}},\label{u-v}
\end{equation}
where 
\begin{equation}
    P_{i \alpha} = \pm \frac{4n_{a}^{0}n_{b}^{0}|V_{ab}|^{2}}{(E_{\mathrm{d}}^{2}-E_{\mathrm{s}}^{2})(E_{\alpha}^{2}-E_{i}^{2})}\label{P_coeff}
\end{equation}
is the positive weight fraction of the $i$th component in the $\alpha$th quasiparticle mode. The upper and lower signs in Eqs.~(\ref{u-v})--(\ref{P_coeff}) correspond, respectively, to the density ($\alpha=\mathrm{d}$) and spin ($\alpha=\mathrm{s}$) modes, and $\zeta=\pm1$ correspond to those coefficients which are conventionally denoted by $u$ and $v$, respectively.

We define the matrix Green functions in the imaginary-time domain as
\begin{equation}
\hat{G}_{ij}(\mathbf{p},\tau)=-\langle T_\tau\left(\begin{array}{c} a_{i\mathbf{p}}(\tau)\\a_{i,-\mathbf{p}}^\dag(\tau)\end{array}\right)\left(\begin{array}{cc}a^\dag_{j\mathbf{p}}(0)&a_{j,-\mathbf{p}}(0)\end{array}\right)\rangle,
\end{equation}
In a two-component Bose-condensed system, these functions can be found from the Dyson-Beliaev equations \cite{Utesov_2018}, and in the frequency domain they can be written as combinations
\begin{equation}
\hat{G}_{ij}(\mathbf{p},i\omega_n)=\sum_{\alpha=\mathrm{d},\mathrm{s}}\sum_{s=\pm}\frac{s}{i\omega_n-sE_\alpha}\left(\begin{array}{c} u_{i\alpha}^s\\u_{i\alpha}^{-s}\end{array}\right)\left(\begin{array}{cc}u_{j\alpha}^s&u_{j\alpha}^{-s}\end{array}\right)\label{Green_functions}
\end{equation}
of positive- and negative-frequency Green functions $1/(i\omega_n\mp E_\alpha)$ of  Bogoliubov quasiparticles weighted with the transformation coefficients (\ref{u-v}).

The density of non-condensate fraction of the $i$th component can be calculated as $n_{i}^{\mathrm{nc}}=A^{-1} \sum_{\mathbf{p}\neq0}\langle a^{\dag}_{i\mathbf{p}} a_{i\mathbf{p}} \rangle$. Using the Bogoliubov transformation (\ref{Transformation}) and taking the thermal averages, we obtain
\begin{equation}
    n_{i}^{\mathrm{nc}}=\frac{1}{A}\sum_{\mathbf{p}}\sum_{\alpha=\mathrm{d,s}}\left\{(u_{i \alpha}^-)^2 +\frac{(u_{i \alpha}^{+})^{2}+(u_{i \alpha}^{-})^{2}}{e^{E_{\alpha}/T}-1}\right\}.\label{non-cond}
\end{equation}

\section{Current response function}\label{app:Details}

We define the Fourier harmonic operator of current as $\mathbf{j}_i(\mathbf{q})=m_{i}^{-1}\sum_{\mathbf{p}}(\mathbf{p}+\frac12\mathbf{q})a_{i\mathbf{p}}^{\dagger}a_{i,\mathbf{p}+\mathbf{q}}$. In the simplest one-loop approximation, which is also used by other authors to describe the conductivity and superfluid drag effect in multi-component ballistic systems \cite{Romito2020, Sekino_2023}, the transverse part of the current response tensor (\ref{chi1}) in Matsubara representation at $q=0$ reads:
\begin{align} 
    \chi^{\mathrm{T}}_{ij}(0,i\omega)&= -T\sum_{\mathbf{p}\omega_{n}}\frac{p^2}{2Adm_{i}m_{j}}\nonumber\\
    &\times\mathrm{Tr}\, [\sigma_z\hat{G}_{ij}(\mathbf{p},i\omega_{n}+i\omega)\sigma_z\hat{G}_{ji}(\mathbf{p},i\omega_{n})].\label{eq120}
\end{align}
Using here the Green functions (\ref{Green_functions}), we obtain Eqs.~(\ref{chiSF})--(\ref{S_clean}) for the current response function. Separating terms with $\alpha_1=\alpha_2$ and $\alpha_1\neq\alpha_2$, we obtain
\begin{equation}
    \chi^{\mathrm{T}}_{ij}(0,i\omega) = \Upsilon_{ij}(i\omega)-\frac{(-1)^{\delta_{ij}}}{m_{i}m_{j}}\left[\Lambda^{+}(i\omega)+\Lambda^{-}(i\omega) \right].\label{chiSF3}
\end{equation}
The function $\Upsilon_{ij}(\omega)$, responsible for intra-branch scattering processes [see Fig.~\ref{fig:Edif}(c,d)], is defined as
\begin{align}
    \Upsilon_{ij}(i\omega)&= \sum_{\mathbf{p}}\sum_{\alpha = \mathrm{d,s}} \frac{p^2P_{i\alpha} P_{j\alpha}}{2Adm_{i}m_{j}}\nonumber\\
    &\times\left[S(E_{\alpha},E_{\alpha})+S(-E_{\alpha},-E_{\alpha})\right].
\end{align}
After introducing the quasiparticle damping and performing analytical continuation $i\omega\rightarrow\omega+i0$, we obtain $S(E_{\alpha},E_{\alpha})=2i\gamma n^{\prime}_{\mathrm{B}}(E_{\alpha})/(\omega+2i\gamma)$ from Eq.~(\ref{approx}), so this function can be written as 
\begin{align}
    \Upsilon_{ij}(\omega)&=\sum_{\mathbf{p}} \frac{i\gamma p^{2}}{Adm_{i}m_{j}}\frac{P_{i\mathrm{d}}P_{j\mathrm{d}}n^{\prime}_{\mathrm{B}}(E_{\mathrm{d}}) + P_{i\mathrm{s}}P_{j\mathrm{s}}n^{\prime}_{\mathrm{B}}(E_{\mathrm{s}})}{\omega+2i\gamma}\nonumber\\
    &=-\frac{2i\gamma D^{0}_{ij}}{\omega+2i\gamma},
\end{align}
where we defined the conductivity weight (\ref{D0_weight}).

The functions $\Lambda^{+}$ and $\Lambda^{-}$ are defined as
\begin{align}
    \Lambda^{\pm}(i\omega)&= \pm \sum_\mathbf{p}\frac{p^{2}}{8Ad}  \sqrt{P_{i\mathrm{d}} P_{i\mathrm{s}} P_{j\mathrm{d}} P_{j\mathrm{s}}} \frac{(E_\mathrm{d}\pm E_\mathrm{s})^2}{E_\mathrm{d}E_\mathrm{s}}\nonumber\\ &\times\sum_{\alpha=\mathrm{d,s}}\left[S(E_{\alpha},\pm E_{\tilde\alpha})+S(-E_{\alpha},\mp E_{\tilde\alpha})\right],
\end{align}
where $\tilde{\mathrm{d}}=\mathrm{s}$ and $\tilde{\mathrm{s}}=\mathrm{d}$. Using the identity $P_{a\mathrm{d}} P_{a\mathrm{s}}=P_{b\mathrm{d}} P_{b\mathrm{s}}$ for the weight factors (\ref{P_coeff}), we obtain for 3D systems the final expression (\ref{app_f}) with the envelope functions
\begin{align}
    &f_{\pm}(p) = \pm \frac{p^4P_{i\mathrm{d}} P_{i\mathrm{s}}}{8\pi^2d} \frac{(E_{\mathrm{d}} \pm E_{\mathrm{s}})^{2}}{E_{\mathrm{d}}E_{\mathrm{s}}}\nonumber\\ 
    &\times\left\{n_{\mathrm{B}}(E_{\mathrm{d}})-n_{\mathrm{B}}(\pm E_{\mathrm{s}}) - i\gamma\left[n^{\prime}_{\mathrm{B}}(E_{\mathrm{d}})+n^{\prime}_{\mathrm{B}}(E_{\mathrm{s}})\right]\right\}.\label{20}
\end{align}
In the Lorentz regime we omit the terms $\gamma n^{\prime}_{\mathrm{B}}$, because $\gamma n^{\prime}_{\mathrm{B}}\sim \gamma / T$, which is much smaller than 1 in realistic systems (since $1\,\mbox{nK}\approx 2\pi\times138\,\mbox{Hz}$, so $T_\mathrm{c}\sim10^2-10^3\,\mbox{nK}$ corresponds to $\sim 10^5\,\mbox{Hz}$). However, these terms should be taken into account in the Drude regime: the conductivity weights (\ref{D0_weight}), (\ref{26}) should be calculated as accurately as possible, because their combined contribution to the intercomponent superfluid weight $D^{\mathrm{s}}_{ab}=\pi(-D^0_{ab}+D^+_{ab}+D^-_{ab})$ can be close to zero due to almost complete canceling of intra- and inter-Bogoliubov branch excitation processes. For instance, for the Rb-Rb mixture with $\gamma/2\pi = 300\,\mbox{Hz}$ and $T=\frac13T_{\mathrm{c}}$, the Drude weights are $D_{ab}^{+}\approx1.02 D_{ab}^{0}$ and $D_{ab}^{-}\approx0.04 D_{ab}^{0}$, and, consequently, $D_{ab}^{\mathrm{s}}\approx0.06\pi D_{ab}^{0}$. Therefore, even small errors in calculations of $D_{ij}^{0,\pm}$ can significantly affect conductivity in the low-frequency limit. 

\section{Approximations for $\Delta$}\label{app:Delta}

The envelope functions (\ref{20}) increase at low momenta, when $E_{\mathrm{d,s}}<T$, due to the power-law factor $p^4$, and then exponentially decrease at large momenta, when $E_{\mathrm{d,s}}>T$, thanks to the Bose distribution functions. Therefore, $f_\pm(p)$ reach maxima near some intermediate momentum $\bar{p}$ where $E_{\mathrm{d,s}}\sim T$. We restrict ourselves to the case when $E_{\mathrm{d}}-E_{\mathrm{s}}\ll E_{\mathrm{d}}+E_{\mathrm{s}}$ and hence $E_\mathrm{d}\approx E_\mathrm{s}$ near $p=\bar{p}$, so that we are able to formally define $\bar{p}$ as a solution of equation $E_{\mathrm{d}}(\bar{p})+E_{\mathrm{s}}(\bar{p})=2T$. In the parameter range we consider, when $T>\mu_{a,b}$, the dispersions $E_{\mathrm{d,s}}$ are almost quadratic near the momentum $\bar{p}$. Using Eq.~(\ref{E_dispersion}) in the this quadratic regime, we can approximate it as $\bar{p}\approx\sqrt{2Tm_{a}m_{b}/(m_{a}+m_{b})}$.

To comprehend behaviour of the integrands in Eq.~(\ref{app_f}), we should consider $E_{\mathrm{d}} + E_{\mathrm{s}}$ and $E_{\mathrm{d}} - E_{\mathrm{s}}$ in denominators of the $R_{\pm}$ functions (\ref{R_pm}) near $p=\bar{p}$. The sum of energies, by definition, is about $E_{\mathrm{d}}(\bar{p})+E_{\mathrm{s}}(\bar{p})=2T$ near this momentum. The difference of energies, denoted as $\Delta=E_{\mathrm{d}}(\bar{p})-E_{\mathrm{s}}(\bar{p})$, can be estimated using quadratic approximation of dispersions (\ref{E_dispersion}): $\Delta \approx \sqrt{(E_a^2-E_b^2)^2 + 16 r^{2}\epsilon_a \epsilon_b \mu_{a}\mu_{b}}/(\epsilon_a + \epsilon_b)$, where $r^{2}=g_{ab}^{2}/g_{aa}g_{bb}$ should be less than 1 for stability of the two-component BEC \cite{Fil2005}. The first term under the square root can be rewritten as $E_a^2-E_b^2 =(\epsilon_{a}^{2} - \epsilon_{b}^{2}) + (2\epsilon_a\mu_{a}-2\epsilon_b\mu_{b})$. It is straightforward to show, that $\Delta\approx|\epsilon_a-\epsilon_b|$ when $|\epsilon_{a}^{2}-\epsilon_{b}^{2}| \gg \epsilon_i\mu_{i}$ at $p=\bar{p}$, which happens when masses of atoms are distant enough: such condition can be written as $|m_{a}-m_{b}|/m_{a}m_{b} \gtrsim \mu_{i}/T m_{i}$. Otherwise, when masses are close to each other, $|\epsilon_{a}^{2} - \epsilon_{b}^{2}|\ll \epsilon_i\mu_i$, we obtain $\Delta\approx\sqrt{\mu^{2}_{a}+\mu^{2}_{b}+2(2-r^{2})\mu_{a}\mu_{b}}\approx\mu_{a}+\mu_{b}$.

Overall, near $p=\bar{p}$, where the envelope functions $f_\pm(p)$ attain the maximum, we obtain the following estimates for sum and difference of the quasiparticle energies:
\begin{equation}
E_{\mathrm{d}} + E_{\mathrm{s}} \sim 2T,\qquad E_{\mathrm{d}} - E_{\mathrm{s}} \sim \Delta,\label{E_pm_E}
\end{equation}
where
\begin{equation}
    \Delta \sim
    \begin{cases} 
    T\dfrac{|m_{a}-m_{b}|}{m_{a}+m_{b}} & \mbox{if}\quad\dfrac{|m_{a}-m_{a}|}{m_{a}m_{b}} \gtrsim \dfrac{\mu_{i}}{Tm_{i}}
    \\[10pt]
    \mu_{a}+\mu_{b} & \mbox{if}\quad\dfrac{|m_{a}-m_{a}|}{m_{a}m_{b}} \ll \dfrac{\mu_{i}}{Tm_{i}}.
    \end{cases}\label{Delta}
\end{equation}
The first and second lines in Eq.~(\ref{Delta}) correspond to the case of distant and close masses, respectively. The symmetric mixtures $m_a=m_b$ are obviously related to the second case. The aforementioned condition $E_{\mathrm{d}}-E_{\mathrm{s}}\ll E_{\mathrm{d}}+E_{\mathrm{s}}$, taken at the most relevant momenta $p\approx\bar{p}$, reduces to $\Delta\ll2T$. In the case of distant masses it reads $|m_{a}-m_{b}|\ll m_{a}+m_{b}$ (thus implying that the mass difference in this case is bounded both above and below), and in the case of close masses it is $\mu_{a}+\mu_{b} \ll T$ (which is fulfilled in the parameter ranges we consider).

\section{Interactions between dipolar atoms}\label{app:d-dint}

In a system of magnetic dipolar atoms, total interatomic interaction 
\begin{equation}\label{Vij_sum}
    V_{ij}(\mathbf{r})=g_{ij}\delta(\mathbf{r})+V_{ij}^{\mathrm{dd}}(\mathbf{r}),
\end{equation}
consists of conventional isotropic interaction due to short-range atomic scattering $g_{ij}\delta(\mathbf{r})$ and long-range magnetic dipole-dipole interaction
\begin{equation}
    V_{ij}^{\mathrm{dd}}(\mathbf{r})=d_{i}d_{j}\frac{1-3\cos^2\theta}{|\mathbf{r}|^{3}},\label{V_dd}
\end{equation}
where $d_{i}$ is a magnetic dipole moment of the $i$th atomic specie, and $\theta$ is the angle between $\mathbf{r}$ and magnetic dipole moments of all atoms which are assumed to be directed along the $z$ axis. 

We consider quasi-two-dimensional atomic clouds with an effective thickness $w_{z}$. In this case 2D Fourier transform of the full intracomponent (i.e. in the same planar cloud) interaction (\ref{Vij_sum}) can be approximated as \cite{Fischer_2006,Boudjem_a_2013,Boudjem_a_2019}
\begin{equation}
    V_{ii}(\mathbf{q})=g_{ii}(1-C_{i} |\mathbf{q}|),
\end{equation}
where $C_{i}=2\pi d_{i}^{2}w_{z}/g_{ii}$ at low enough momenta (full momentum dependence was analyzed in Ref.~\cite{Fischer_2006}). This interaction potential is evaluated with assumption $r_*q\ll1$, where $r_{*}=m_{i}d_{i}^{2}$ is the characteristic range of dipole-dipole interaction. This assumption is valid for the parameters used in our calculations: $r_{*}=12\,\mbox{nm}$ and $2.6\,\mbox{nm}$ for $^{168}$Er and $^{52}$Cr atoms respectively is much smaller than interlayer distance $L=60\,\mbox{nm}$, which determines the scale of inverse momentum $q^{-1}$.

The Fourier transform of interaction between particles in different spatially separated atomic clouds is found as follows. First, we rewrite the interaction (\ref{V_dd}) for the two-layer geometry: 
\begin{align}
    V_{ab}^{\mathrm{dd}}(\mathbf{\rho},z-z^{\prime})&=d_{a}d_{b}\frac{r^2-3(L+z-z^{\prime})^2}{r^5}\nonumber\\
    &\times \left(\frac{\pi}{2w_{z}}\right)^{2}\cos\left(\frac{\pi z}{w_{z}}\right)\cos\left(\frac{\pi z^{\prime}}{w_{z}}\right),\label{V_ab_dipole_coord}
\end{align}
where $\rho$ and $L+z-z^{\prime}$ are in-plane and out-of-plane distances between two atoms, while $r=\sqrt{(L+z-z^{\prime})^{2}+\rho^2}$ is the total distance; $z$ and $z^{\prime}$ are their vertical coordinates relative to the cloud centers ranging from $-w_{z}/2$ to $w_{z}/2$. The cosine functions model atomic density profiles in the $z$-axis direction, and $(\pi / 2 w_{z} )^{2}$ is normalization factor. 2D Fourier transform of Eq.~(\ref{V_ab_dipole_coord}) in the $xy$ plane reads
\begin{align}
    V_{ab}^{\mathrm{dd}}(\mathbf{q},z,z^{\prime})&= -2\pi d_{a}d_{b} qe^{-q(L+z-z^{\prime})}\nonumber\\ 
    &\times \left(\frac{\pi}{2w_{z}}\right)^{2} \cos\left(\frac{\pi z}{w_{z}}\right)\cos\left(\frac{\pi z^{\prime}}{w_{z}}\right).\label{V_dd_2D}
\end{align}
We assume thinness of atomic clouds,  $w_{z}\ll L$ (which was achieved in the recent experiment \cite{DuLi_2023}), so out-of-plane momenta of interacting particles are almost unchanged by the interlayer interaction. Therefore we can integrate the interaction (\ref{V_dd_2D}) over $z$ and $z^{\prime}$, arriving at the formula
\begin{equation}
    V_{ab}^{\mathrm{dd}}(\mathbf{q})=-2\pi d_{a}d_{b} q e^{-qL}\left[ \frac{\cosh(qw_{z}/2)}{1+q^{2}w_{z}^{2}/\pi^{2}}\right]^{2},\label{dd_3D}
\end{equation}
which is used in numerical calculations for Fig.~\ref{fig:tr-dd}. The two-dimensional condensate density of each component is estimated as $n_{i}^{0}w_{z}$, where the typical three-dimensional densities $n_{i}=\zeta (3/2) \left(m_{i}T_\mathrm{c}^i/2\pi\right)^{3/2}$ are related to the critical temperatures $T_\mathrm{c}^i$ taken from Table~\ref{table:1}. 

\bibliography{References.bib}

\end{document}